\documentclass[aps,prd,amsmath,twocolumn,showpacs]{revtex4}

\usepackage{epsfig}
\usepackage{graphics}
\usepackage{latexsym}
\usepackage{amsmath}
\usepackage{amssymb}
\usepackage{rotating}
\usepackage{subfigure}
\usepackage{bm}
\usepackage{color}
\usepackage{ulem}
\usepackage{booktabs}

\usepackage[colorlinks = true,
linkcolor = magenta,
urlcolor  = blue,
citecolor = red,
anchorcolor = blue]{hyperref}

\begin{document}

%
%
\title{ First global QCD analysis of charged hadron fragmentation functions and their uncertainties at next-to-next-to-leading order }
%
%

\author{Maryam Soleymaninia$^{1,3}$}
\email{Maryam\_Soleymaninia@ipm.ir}

\author{Muhammad Goharipour$^{3}$}
\email{Muhammad.Goharipour@ipm.ir}

\author{Hamzeh Khanpour$^{2,3}$}
\email{Hamzeh.Khanpour@mail.ipm.ir}

\affiliation {
$^{1}$Department of Physics, Shahid Rajaee Teacher Training University, Lavizan, Tehran 16788, Iran          \\
$^{2}$Department of Physics, University of Science and Technology of Mazandaran, P.O.Box 48518-78195, Behshahr, Iran \\
$^{3}$School of Particles and Accelerators, Institute for Research in Fundamental Sciences (IPM), P.O.Box 19395-5531, Tehran, Iran       
 }

\date{\today}

%
\begin{abstract}

In this paper, we present ${\tt SGK18}$ FFs, a first global QCD analysis of parton-to-{\it unidentified} charged hadrons fragmentation functions (FFs) at next-to-next-to-leading order (NNLO) accuracy in perturbative QCD. This analysis is based on single-inclusive charged hadron production in electron-positron ($e^-e^+$) annihilation. The uncertainties in the extraction of ${\tt SGK18}$ FFs as well as the corresponding observables are estimated using the ``Hessian'' technique. We study the quality of the {\tt SGK18} FFs determined in this analysis by comparing with the recent results in literature. We also show how {\tt SGK18} FFs results describe the available data for single-inclusive unidentified charged hadron production in  $e^-e^+$ annihilation. We demonstrate that the theoretical uncertainties due to the variation of the renormalization and factorization scales improve when NNLO QCD corrections are considered. We find that the resulting {\tt SGK18} FFs are in good agreement with all data analyzed and the inclusion of NNLO corrections tends to improve the data description with somewhat smaller uncertainty.

\end{abstract}

\maketitle
\tableofcontents{}

%
\section{Introduction}\label{sec:introduction}
%

Parton distribution functions (PDFs) extracted from global QCD fits of deep inelastic electron proton ($ep$) scattering at HERA as well as proton-(anti)proton ($pp$) collision at hadron colliders such as LHC, are a fundamental input into hadron collider physics and has been very important in the investigation of the partonic structure of the nucleon, (see~\cite{Abt:2016ttq,Lin:2017snn,Harland-Lang:2017xtj,Ball:2018odr} for a recent review and ~\cite{Harland-Lang:2017ytb,Ball:2017nwa,Bourrely:2015kla,Harland-Lang:2014zoa,Hou:2017khm,Alekhin:2017kpj,Khanpour:2016pph,Eskola:2016oht,Kovarik:2015cma,Wang:2016mzo,Khanpour:2017cha,Jimenez-Delgado:2014xza,Sato:2016tuz,Nocera:2014gqa,Ethier:2017zbq,Khanpour:2017fey,Goharipour:2018yov} for recent determination of different types of PDFs). 
Much effort also have been made by theoretical and experimental particle physics communities to improve our understanding on the partonic structure of the nucleon and nuclei~\cite{Lin:2017snn,Gao:2017yyd,Ball:2018iqk,South:2016cmx,Ceccopieri:2016rga,Ceccopieri:2018nop,Selyugin:2017kwa,Selyugin:2014sca,Selyugin:2015pha,Britzger:2018zvv,Boroun:2018lpz,Zarrin:2016kxf,Frankfurt:2016qca,Guzey:2016tek,Lalung:2018mpw,Arbabifar:2013tma,Rasmussen:2018dgo,Zarei:2015jvh,MoosaviNejad:2016qdx,MoosaviNejad:2017rvi,MoosaviNejad:2016jpc}. These studies mostly include the the potential of recent measurements at high energy collider to better constrain our present knowledge of the PDFs and the importance or resulting PDFs for predictions of processes at the LHC and possible future high energy and high luminosity lepton and hadron colliders.

Like PDFs, fragmentation functions (FFs) also plays an important role in our understanding of certain high energy processes with identified hadrons in the final state~\cite{Bertone:2017tyb}. 
According to the asymptotic freedom of QCD, fragmentation functions (FFs) relates to the long-distance dynamics of the interactions among quarks and gluons which cause to their hadronization in a hard-scattering process~\cite{Field:1977fa,Collins:1981uk}. The experimental observables of single-inclusive hadron production involve identified hadrons in the final state. In order to obtain theoretical predictions for such processes, it is written down a factorization formula and FFs are convoluted with partonic cross-sections.
FFs plays an important role in understanding of non-perturbative QCD dynamics. Like for the case of PDFs, FFs are determined from global QCD analysis of experimental measurements particularly in hadronization processes. 
These processes include single-inclusive hadron production of electron-positron ($e^+e^-$) annihilation (SIA), semi-inclusive deep-inelastic lepton-nucleon ($\ell^\pm {\rm N}$) scattering (SIDIS), and single-inclusive hadron production in proton-proton ($pp$) collisions.
According to the QCD-improved parton model, FFs and PDFs scaling violations are subjects to the perturbatively computable Dokshitzer-Gribov-Lipatov-Alteralli-Parisi (DGLAP) evolution equations~\cite{Gribov:1972ri,Lipatov:1974qm,Altarelli:1977zs,Dokshitzer:1977sg}.

In the past few years, several progresses have been done to determine FFs for light and heavy mesons which performed at next-to-leading order (NLO) and next-to-next-to-leading order (NNLO) accuracy in perturbative QCD~\cite{Bertone:2017tyb,Anderle:2015lqa,Soleymaninia:2017xhc}. At NNLO, only experimental data
from electron-positron annihilation can be used in a QCD analysis while the calculations for the hard processes in SIDIS and $pp$ collisions at NNLO are not accessible yet. Electron-positron annihilation provides the most cleanest and appropriate data sets to access to the FFs, where the final state quarks and gluons fragment into hadrons. Compared to the SIDIS and $pp$ collision, the FFs in SIA processes are the only non-perturbative functions in the calculation of the cross section. Since the SIA experimental measurements are not sensitive to the separation between quark and antiquark FFs, the extraction of favored and disfavored fragmentations is difficult in this processes. In order to allow one to determine quark from antiquark FFs, the data where hadrons of different electrical charge are identified in the final state needs to be taken into account~\cite{deFlorian:2007aj}. 
In addition, the gluon fragmentation density is not exceedingly well constrained by SIA data, since the subleading NLO and NNLO corrections for $e^- e^+$ annihilation are too weak to determine it. Including the data from SIDIS and $pp$ collision in the extraction of FFs could increase the statistics and also provide a much more complete picture of the fragmentation processes. In {\tt SGK18}  FFs we restrict our analysis to the SIA data, since the QCD framework for FFs at NNLO only accessible for the $e^- e^+$ annihilations.  

Historically, the knowledge of FFs and their determination through the global analysis of the experimental data has undergone many developments both experimentally and theoretically. For example,
in the analysis of Ref.~\cite{Soleymaninia:2013cxa}, the authors used simultaneously the SIA and SIDIS asymmetry data from the {\tt HERMES}~\cite{Airapetian:2004zf} experiment at HERA and {\tt COMPASS}~\cite{Alekseev:2009ac,Alekseev:2010ub} experiment at CERN to determined the pion and kaon FFs both at leading order (LO) and NLO approximation. In the analysis of Ref.~\cite{Nejad:2015fdh}, the authors considered the finite-mass effects of the proton to calculate the proton FFs by including SIA data at LO and NLO accuracies. Recently, pion, kaon and proton FFs have been extracted by various groups such as the {\tt DEHSS}~\cite{deFlorian:2014xna,deFlorian:2017lwf}, {\tt HKKS}~\cite{Hirai:2016loo}, {\tt JAM}~\cite{Sato:2016wqj}, and also by the {\tt NNPDF} Collaboration~\cite{Bertone:2017tyb} using the iterative Monte Carlo method. For the case of charmed-meson $D^*$ FFs, we refer the reader to the very recent {\tt AKSRV17}~\cite{Anderle:2017cgl} and {\tt SKM18}~\cite{Soleymaninia:2017xhc} global analyses. It should be note that the later one has been done at for the first time at NNLO approximation by including the SIA data.

In this paper, we perform for the first time a comprehensive QCD analysis to obtain a set of  {\it unidentified} charged hadron FFs and their uncertainties at NNLO. In order to perform our global analysis for determining the FFs of the {\it unidentified} charged hadrons at NNLO, we have to limit the potential of global
determination of FFs to the SIA measurements. We show that the inclusion of higher order QCD correction could describe the data well, including those data points at rather smaller hadron momentum fraction $z$, $z < 0.02$. We extensively discuss the theoretical and phenomenological methodology of the {\tt SGK18} analysis, including the exprimental description of the $e^- e^+$ annihilation exprimental observables in term of the {\tt SGK18} FFs, parameterizations and the fitting procedure in next sections of this paper. 

Previously available analyses of inclusive charged hadron FFs sum up the pion, kaon and (anti)proton results and ignore the contributions of possible heavier charged hadrons. For example, in Ref.~\cite{Albino:2005me}, the inclusive charged hadron experimental data have been excluded from the analysis and only the sum of charged pion, kaon and protons FFs obtained from the fit has been compared to the inclusive charged hadron data. While in Ref.~\cite{Bourhis:2000gs}, the FFs for {\it unidentified} charged hadrons has been extracted. 
The NNPDF Collaboration, after having extracted trustworthy FFs for pion, kaon and proton, as the lightest and most copiously produced charged hadrons, has recently calculated the FFs for {\it unidentified} charged hadron up to NLO accuracy~\cite{Nocera:2017gbk}. In addition, the analysis by {\tt DSS07}~\cite{deFlorian:2007ekg} included the electron-proton annihilations, SIDIS and proton-(anti)proton collisions experimental data sets. They obtained the contributions from the residual charged hadrons as well as pions, kaons and (anti)protons to the {\it unidentified} charged hadron FFs up to NLO accuracy. In this work, we extend the extraction of FFs for charged hadrons for the first time up to NNLO by including the inclusive charged hadron experimental data from $e^-e^+$ annihilation.

In order to assess the uncertainties of the resulting {\it unidentified} charged hadron FFs at NLO and NNLO accuracies as well as the corresponding observable, associated with the uncertanties in the analyzed data, we have applied the ``Hessian'' method. 

This paper is organized as follows:
In Sec.~\ref{sec:data selection}, the datasets included in {\tt SGK18} FFs analysis, along with the corresponding observables and kinematic cuts are presented.
We discuss the QCD analysis of hadronization process in electron-positron annihilation by introducing FFs and their evolution in Sec.~\ref{sec:QCD analysis}.
We describe our formalism, input parametrization at the initial scale for the determination of {\it unidentified} charged hadron FFs in Sec.~\ref{sec:parametrization}.
In Sec.~\ref{sec:errorcalculation}, the minimization strategy and the ``Hessian'' uncertainty approach to calculate the errors of {\tt SGK18} FFs analysis are presented.
In Sec.~\ref{sec:Results}, we present the obtained results for the $D^h$-FFs and their uncertainties. We also perform a comparison of {\tt SGK18} results with the analyzed experimental data and other available FF sets in this section. 
The theoretical uncertainties, fit quality and the stability due to the variation of the renormalization and factorization scales are studied at the end of this section. 
Finally, we conclude and summarize the results in Sec.~\ref{sec:conclusion}.

\section{ Experimental observables } \label{sec:data selection}

We begin this section with discussing the measurements of charged hadron production in $e^+ e^-$ annihilation, collected by a variety of experiments~\cite{Buskulic:1995aw,Ackerstaff:1998hz,Akers:1995wt,Abreu:1998vq,Abreu:1997ir,Aihara:1988su,Abe:2003iy,Braunschweig:1990yd} at CERN, SLAC and HERA.
Our aim is to include all available data sets which help to constrain the resulting charged hadron FFs, and more importantly, provide additional consistency checks of the fitting procedure. 

In this analysis, the FFs are determined by including a wide range of the experimental data from electron-positron annihilation into an {\it unidentified} charged hadron {\it {h}} and the unobserved jets which are produced along with the detected hadron $h$. This process is given by:

\begin{eqnarray}\label{SIA}
e^{+} + e^{-} \rightarrow (\gamma, Z^0) \rightarrow {\it h} + X \,.
\end{eqnarray}

The DIS process is space-like, while the above process is time-like and the related scaling variable is $z = 2 p_h.q/Q^2$, in which 
 the four-momenta of the intermediate gauge boson and hadron $h$ have been denoted by $q$ and $p_h$, respectively, with $\sqrt{q^2} = Q$. 
 In the center-of-mass energy frame where $\sqrt{s}=Q$, the scaling variable can be written as $z=2E_h/{\sqrt{s}}$.

In this analysis, the analyzed data sets are based on SIA differential cross-sections for the {\it unidentified} charged hadron $h=h^+ + h^-$.
These data sets are differential with respect to the scaling variable $z$ or $p_h$. Actually, the format of the experimental data are different among the various experiments. In Table.~\ref{tab:datasets}, the SIA cross sections included in {\tt SGK18} analysis have been listed for different experiments. The kinematical variables are as follows:  scaling variable $z = 2E_h/\sqrt{s}$, the observed hadron $h$ energy that scaled to the beam energy, and the hadron three-momentum $p_h$. The scaled momentum $x_p$ is given by $x_p=2p_h/\sqrt{s}$. The relation between scaled momentum $x_p$ and $z$ is defined as

\begin{equation}\label{xp}
z = \sqrt{(1 - \rho_{\it h}) x_p^2 + \rho_{\it h}},
\end{equation}
%
where $\rho_{\it h} = 4m^2_{\it h}/s$ and $m_{\it h}$ stands for the hadron mass. Note that ignoring the hadron mass leads to $z = x_p$.

In Table~\ref{tab:datasets} we have listed all analyzed flavor-untagged and tagged measurements used in our analysis which are reported by different experiments. These data sets include the {\tt ALEPH}~\cite{Buskulic:1995aw}, {\tt OPAL}~\cite{Ackerstaff:1998hz,Akers:1995wt} and {\tt DELPHI}~\cite{Abreu:1998vq,Abreu:1997ir} experiments at CERN; the {\tt TPC}~\cite{Aihara:1988su} and {\tt SLD}~\cite{Abe:2003iy} experiments at SLAC; and {\tt TASSO}~\cite{Braunschweig:1990yd} experiment at DESY.
As one can see from Table~\ref{tab:datasets}, the measured observables are different for these data sets. Most of experimental collaborations have reported total inclusive and tagged cross sections, while the {\tt ALEPH}, {\tt DELPHI} and {\tt OPAL} have reported longitudinal inclusive and bottom tagged cross section data. 
Separation of light and heavy quark flavor FFs is provided by the light and heavy flavor tagged experimental data.
The longitudinal cross section data are proportional to the longitudinal structure function $F_L$ and implemented in {\tt SGK18} analysis to put further constraints on the gluon fragmentation function. The gluon coefficient functions were already available from several years ago at LO ${\cal O}(\alpha _s)$. The NLO ${\cal O}(\alpha ^{2}_s)$ coefficient functions have been also used in several analyses, for example in Refs.~\cite{Rijken:1996ns,Mitov:2006wy}.  However, there is no analysis to determine the FFs of the {\it unidentified} charged hadrons including the coefficient functions at NNLO. As we mentioned the determination of {\it unidentified} charged hadrons at NNLO is the aim of the present paper.

Another point should be mentioned here is on the kinematic cuts applied on the data sets in {\tt SGK18} FFs analysis. We study the SIA data in potentially problematic low-$z$ region, and hence, kinematic cuts are chosen consistently. To be on the safe side, we exclude the data points below the scaling variable of $z_{min} = 0.02$ for the data sets at $\sqrt s = M_Z$, and $z_{min}=0.075$ for $\sqrt s<M_Z$. The data points with $z_{max} = 0.9$ are not included in {\tt SGK18} QCD fit. The number of data points which are included in {\tt SGK18} fits are shown in the fifth column of Table ~\ref{tab:datasets} for each data sets separately.
Moreover, the quality of our fits to SIA data for {\it unidentified} charged hadron at NLO and NNLO accuracy in term of the individual $\chi^2$ values for every data set are also reported in the last two columns. The total $\chi^2/{\rm d.o.f}$ obtained from {\tt SGK18} best fits can also be found at the bottom of this table which are equal to $1.64$ and $1.62$ for NLO and NNLO analyses, respectively.
Using the total 474 data points, we determine the 20 free parameters describing ${\tt SGK18}$ {\it unidentified} charged hadron FFs $D^h_i(z, Q_0^2)$. The details of ${\tt SGK18}$ analysis on {\it unidentified} charged hadron FFs at NLO and NNLO will be discussed in details in Sec.~\ref{sec:parametrization}.

%
\begin{table*}[htb]
	\renewcommand{\arraystretch}{2}
	\centering 	\scriptsize
	\begin{tabular}{lc||l|cl||clc}
		\toprule \hline \hline
		{\tt Experiment} ~&~ {\tt Reference} ~&~ {\tt Observable} ~&~  $\sqrt{s}$~[{\tt  GeV}] ~&~  {\tt Number of data points} ~&~ $\chi^2$ ({\tt  NLO}) ~&~ $\chi^2$ ({\tt NNLO}) \\ \hline \hline
		\midrule 
		{\tt TASSO22} & \cite{Braunschweig:1990yd}
		& $\frac{1} {\sigma_{\rm{total}}}~\frac{d \sigma^{{\it h}^\pm}} {dz}$ 		& 22.00  		& ~~~~~15	& 8.22& 10.38   \\
		{\tt TASSO14} & \cite{Braunschweig:1990yd} & $\frac{1} {\sigma_{\rm{total}}}~\frac{d\sigma^{{\it h}^\pm}} {dz}$ 		& 14.00  		& ~~~~~15  		& 17.32 & 23.46\\
		{\tt TASSO35} & \cite{Braunschweig:1990yd}
		& $\frac{1} {\sigma_{\rm{total}}}~\frac{d\sigma^{{\it h}^\pm}} {dz}$ 		& 35.00  		& ~~~~~15 	& 14.27& 24.35   \\ 
		{\tt TASSO44} & \cite{Braunschweig:1990yd}
		& $\frac{1} {\sigma_{\rm{total}}}~\frac{d\sigma^{{\it h}^\pm}} {dz}$ 		& 44.00  		& ~~~~~15  		& 8.79& 9.97   \\
		{\tt TPC} & \cite{Aihara:1988su}
		& $\frac{1} {\sigma_{\rm{total}}}~\frac{d\sigma^{{\it h}^\pm}} {dz}$ 		& 29.00  		& ~~~~~21 		& 21.40&  38.67   \\
		 		{\tt ALEPH} & \cite{Buskulic:1995aw}
		& $\frac{1} {\sigma_{\rm{total}}}~\frac{d\sigma^{{\it h}^\pm}} {dz}$ 		& 91.20  		& ~~~~~32  		& 79.91 &90.06   \\
		 		& \cite{Buskulic:1995aw}
		& $\frac{1} {\sigma_{\rm{total}}}~\frac{d\sigma^{{\it h}^\pm}_L} {dz}$ 		& 91.20  		& ~~~~~19  		& 60.64& 12.88 \\ 
				{\tt DELPHI} & \cite{Abreu:1998vq}   
		& $\frac{1} {\sigma_{\rm{total}}}~\frac{d\sigma^{{\it h}^\pm}}{dp_{\it h}}$ 		& 91.20  		& ~~~~~22  		& 31.13& 28.29  \\
		 		& \cite{Abreu:1998vq}
		& $\left.\frac{1} {\sigma_{\rm{total}}}~\frac{d\sigma^{{\it h}^\pm}}{dp_{\it h}}\right |_{uds}$ 		& 91.20  		& ~~~~~22  		& 14.05& 14.46  \\
		 		& \cite{Abreu:1998vq}
		& $\left.\frac{1} {\sigma_{\rm{total}}}~\frac{d\sigma^{{\it h}^\pm}}{dp_{\it h}}\right |_{b}$ 		& 91.20  		& ~~~~~22  		& 60.75& 63.68  \\
		 		& \cite{Abreu:1997ir}
		& $\frac{1} {\sigma_{\rm{total}}}~\frac{d\sigma^{{\it h}^\pm}_L} {dz}$ 		& 91.20  		& ~~~~~20  		& 41.48 & 9.01   \\ 
				& \cite{Abreu:1997ir}
		& $\left.\frac{1} {\sigma_{\rm{total}}}~\frac{d\sigma^{{\it h}^\pm}_L} {dz}\right |_{b}$ 		& 91.20  		& ~~~~~20 		& 9.90&  9.43  \\ 
				{\tt OPAL} & \cite{Ackerstaff:1998hz}
		& $\frac{1} {\sigma_{\rm{tot}}}~\frac{d\sigma^{{\it h}^\pm}} {dz}$ 		& 91.20  		& ~~~~~20   		& 47.49 &  47.78  \\ 
				& \cite{Ackerstaff:1998hz}
		& $\left.\frac{1} {\sigma_{\rm{total}}}~\frac{d\sigma^{{\it h}^\pm}} {dz}\right |_{uds}$ 		& 91.20  		& ~~~~~20   		& 19.97& 19.26  \\ 
				& \cite{Ackerstaff:1998hz}
		& $\left.\frac{1} {\sigma_{\rm{total}}}~\frac{d\sigma^{{\it h}^\pm}} {dz}\right |_{c}$ 		& 91.20  		& ~~~~~20  		& 14.49&  15.95  \\ 
				& \cite{Ackerstaff:1998hz}
		& $\left.\frac{1} {\sigma_{\rm{total}}}~\frac{d\sigma^{{\it h}^\pm}} {dz}\right |_{b}$ 		& 91.20  		& ~~~~~20  		& 18.97&   23.23  \\ 
				& \cite{Akers:1995wt}
		& $\frac{1} {\sigma_{\rm{total}}}~\frac{d\sigma^{{\it h}^\pm}_L} {dz}$ 		& 91.20  		& ~~~~~20  		& 13.99 &  8.60  \\ 
				{\tt SLD} & \cite{Abe:2003iy}
		& $\frac{1} {\sigma_{\rm{total}}}~\frac{d\sigma^{{\it h}^\pm}}{dp_h}$ 		& 91.28  		& ~~~~~34 		& 35.82 &  33.30  \\ 
				& \cite{Abe:2003iy}
		& $\left. \frac{1} {\sigma_{\rm{total}}}~\frac{d\sigma^{{\it h}^\pm}}{dz} \right |_{uds}$ 		& 91.28   		& ~~~~~34 		& 58.55 & 58.80  \\ 
				& \cite{Abe:2003iy}
		& $\left. \frac{1} {\sigma_{\rm{total}}}~\frac{d\sigma^{{\it h}^\pm}}{dz} \right |_{c}$ 		& 91.28   		& ~~~~~34  		& 40.35 & 61.33  \\ 
				& \cite{Abe:2003iy}
		& $\left. \frac{1} {\sigma_{\rm{total}}}~\frac{d\sigma^{{\it h}^\pm}}{dz} \right |_{b}$ 		& 91.28  		& ~~~~~34 		& 128.73 &  136.40  \\ \hline \hline
		\midrule
		{\bf Total data}       &         &       &         &      ~~~~~474  &      746.22 &    739.29 \\
		$\chi^2/{\rm d.o.f}$      &         &       &         &     &      1.64 &1.62 \\ \hline \hline
		\bottomrule
	\end{tabular}
	\caption{ \small The data sets included in {\tt SGK18} analysis of FFs for {\it unidentified} charged hadrons. 
		For each experiment, we indicate the corresponding reference, the measured 
		observables, the center-of-mass energy $\sqrt{s}$, the number of data points 
		included after (before) kinematic cuts, the $\chi^2$ for every data set, and the total $\chi^2/{\rm d.o.f}$. The details
		of corrections to data sets and the kinematic cuts applied are contained in the text. }
	\label{tab:datasets}
\end{table*}

%
%
\section{ The QCD Framework of {\tt SGK18} FFs analysis }\label{sec:QCD analysis}
%

In the present {\tt SGK18} FFs analysis, we work in the well established pQCD framework for the electron-positron SIA process at the NLO and NNLO accuracy in pQCD. We make an extensive use of the $x$-space DGLAP evolution implemented in publicly available {\tt APFEL} code~\cite{Bertone:2013vaa} in which developed for a fast computation of the NLO and NNLO cross section of $e^- e^+$ annihilation. For a clear review, we refer the reader to the Ref.~~\cite{deFlorian:2007aj,Bertone:2013vaa,Stratmann:2001pb} for further technical details of the QCD framework. 

In this section, we review the factorization theorem of the cross-section and fragmentation structure functions in the electron-positron SIA process. We also discuss the time-like DGLAP evolution of FFs. The differential cross-section for the single-inclusive $e^+ e^-$ annihilation involving a hadron $h$ in the final state,

\begin{equation}
e^+ e^- \rightarrow (\gamma, \,Z) \rightarrow {\it h} \,,
\end{equation}
%
with integrated over the production angle, and at a center-of-mass framework energy of $s$, is given by:

\begin{equation}\label{diffcross}
\frac{1}{\sigma_{\rm tot}}
\frac{d \sigma^{\it h}} {dz} =
\frac{1} {\sigma_{\rm tot}}
\left[ F_T^{\it h}(z, Q^2) +
F_L^{\it h}(z, Q^2) \right] \,,
\end{equation}
%
where $ F_T^{\it h}(z, Q^2) $ and $ F_L^{\it h}(z, Q^2) $ are the transverse and longitudinal structure functions, respectively.

In the case of multiplicities, the total cross section for the electron positron annihilation into hadrons normalized to the differential 
cross section up to NNLO is written as: 

\begin{equation}\label{sigmatot}
\sigma_{\rm tot} =
\sum_q \hat{e}_q^2\; \sigma_0
\left[1+ \alpha_sK^{(1)}_{\rm QCD} +
\alpha_s^2K^{(2)}_{\rm QCD}
 + ...\right] \,,
\end{equation}
%
where the coefficients $K^{(i)}_{\rm QCD}$ relate to the QCD perturbative corrections that are currently known up to ${\cal O}(\alpha _s^3)$~\cite{Gorishnii:1990vf}.
Note that we have integrated over the scattering angle $\theta$ of the hadron $h$, and the cross section can be decomposed into transverse (T) and  longitudinal (L) parts.
Then $F_T^{h}$ and $F_L^{h}$ are called the time-like structure functions or fragmentation structure functions.
The NNLO QCD corrections to the fragmentation structure functions can be expressed in factorized form of fragmentation functions $D^{\it h}_{i}(z, Q^2)$ and
calculable coefficient functions $C^{\rm S, NS}_{k, l}(z, \alpha _s(Q))$ as follows

\begin{eqnarray}\label{structure functions}
F_k^{\it h}(z, Q^2)&=& \sigma_{\rm tot}^{(0)}[D^{\it h}_S(z, Q^2)
\otimes C^S_{k,q}(z, \alpha _s(Q)) \nonumber \\
&+& D^h_g(z, Q^2) \otimes C^S_{k, g}(z, \alpha _s(Q))] \nonumber \\
&+& \Sigma_{q} \sigma ^{(0)}_q D^{\it h}_{\rm NS,q}(z, Q^2)
\otimes C^{\rm NS}_{k, q}(z, \alpha _s(Q))\,. \nonumber\\
\end{eqnarray}

The coefficient functions $C^{\rm S, NS}_{k, l}$ with $k=T, L$ and $l=q, g$ have been calculated in Refs.~\cite{Rijken:1996ns, Mitov:2006wy, Rijken:1996vr}.
The factorization scale $\mu _F$ and the renormalization scale $\mu _R$ are set to be equal to the center-of-mass energy of the collision, $\mu _F = \mu _R = \sqrt{s} = Q$.
In Eq.~\ref{structure functions}, $\sigma_q^{(0)}$ is the total cross section for quark production $q$ at LO and $\sigma_{tot}^{(0)}$ is the corresponding sum over all active flavors $n_f$, $\sigma ^{(0)}_{\rm tot} = \Sigma _q \sigma ^{(0)}_q$.
In this equation, symbol $\otimes$ also denotes the standard convolution integral defined as 

\begin{equation}\label{convolution}
f(z) \otimes g(z)
=\int^1_0 dx \int^1_0 dy f(x) g(y) \delta(z - xy) \,.
\end{equation}

The FFs, $D^{\it h}_i(z,Q^2)$, which are non-perturbative but universal functions, parametrize the hadronization of massless partons, $i=q, \bar{q}, g$, into the observed hadron $h$ which carry  fraction $z$ of the hadron momentum.
The scale dependence of the FFs which are governed by the renormalization equations are calculable in pQCD using the DGLAP evolution equation.
The quark singlet (S) FF $D^{\it h}_{\rm S}(z, Q^2)$, non-singlet (NS) FFs $D^{\it h}_{\rm NS}(z, Q^2)$ as well as the gluon-to-hadron FF $D^h_g(z, Q^2)$ are used in Eq.~\ref{structure functions}, and the singlet and non-singlet FFs are defined as: 

\begin{eqnarray}\label{singlet}
D_S^{\it h}(z, Q^2) = \frac{1}{n_f}
\Sigma_q [D^{\it h}_q(z, Q^2) + D^{\it h}_{\bar{q}}(z, Q^2)] \,,
\end{eqnarray}
%
and 
%
\begin{eqnarray}\label{nonsinglet}
D_{\rm NS, q}^{\it h}(z, Q^2) =
D^{\it h}_q(z, Q^2) +
D^{\it h}_{\bar{q}}(z, Q^2) - D^{\it h}_{\rm S}(z, Q^2).
\end{eqnarray}

The DGLAP evolution equations \cite{Gribov:1972ri,Lipatov:1974qm,Altarelli:1977zs,Dokshitzer:1977sg} evaluate the FFs with the energy scale $Q^2$ as

\begin{eqnarray}\label{DGLAP}
\frac{\partial}{\partial 
\ln Q^2} D_i^{\it h}(z, Q^2) =
\Sigma_j P_{ji}(z, \alpha_s(Q))
\otimes D^{\it h}_j(z, Q^2),
\end{eqnarray}
%
where $i,j=q, \bar{q}, g$ and $P_{ji}$ are the time-like splitting functions~\cite{Mitov:2006ic,Moch:2007tx,Almasy:2011eq}.
According to the different FFs as non-singlet, singlet and gluon FFs, one can rewrite Eq.~\ref{DGLAP} as a decoupled DGLAP equation

\begin{eqnarray}\label{non-singlet DGLAP}
\frac{\partial}
{\partial \ln Q^2} D_{\rm NS}^{\it h}(z, Q^2) =
P^+(z, \alpha_s(Q)) 
\otimes D^{\it h}_{\rm NS}(z, Q^2),
\end{eqnarray}
%
for the non-singlet FFs and two coupled equations for the singlet and gluon FFs as

\begin{eqnarray}\label{singlet DGLAP}
&&\frac{\partial}
{\partial \ln Q^2}
\left(\begin{array}{ccc}
D_S^{\it h} \\
D_g^{\it h} \\
\end{array} \right)(z, Q^2)=
\left(\begin{array}{cccc}
P_{qq} \quad   2n_fP_{gq}\\
\frac{1}{2n_f} P_{qg}  \quad  P_{gg}\\
\end{array} \right) \nonumber \\
&&(z, \alpha_s(Q)) \otimes
\left(\begin{array}{ccc}
D_S^{\it h} \\
D_g^{\it h} \\
\end{array}\right)(z, Q^2).
\end{eqnarray}

The coefficient functions in Eq.~\ref{structure functions} and the splitting functions in Eqs.~\ref{non-singlet DGLAP} and \ref{singlet DGLAP} are defined as a perturbative expansion in powers of the $\alpha_s$,

\begin{eqnarray}\label{wilsons}
C^{\rm S, NS}_{k, i}(z, \alpha _s) =
\Sigma_{l=0}a^l_s C^{\rm S, NS(l)}_{k, i}(z), \nonumber \\
P^{ji, +}(z, \alpha_s) =
\Sigma_{l = 0} a^{l + 1}_s P^{ji, +(l)}(z),
\end{eqnarray}
%
where $i,j=q,g$; $k=T,L$ and $a_s = \alpha_s/(4\pi)$.
In the $\overline{\rm MS}$ scheme, the SIA coefficient functions have been computed up to NNLO for the $C^{\rm S,NS}_{T,i}$.
The longitudinal coefficient functions $C^{\rm S,NS}_{L,i}$ vanish at ${\cal O}(a_s^0)$ and have been reported up to NLO accuracy in Refs.~\cite{Rijken:1996ns,Mitov:2006wy,Rijken:1996vr,Rijken:1996npa,Blumlein:2006rr}.
We should note here that, since $C_{T,g}^{i,(0)}=0$, the gluon FF does not have contribution directly to the LO in SIA case.  
The time-like splitting functions have been calculated up to ${\cal O}(a_s^3)(k=2)$ and can be found in Refs.~\cite{Mitov:2006ic,Moch:2007tx,Almasy:2011eq}.

Our aim in this analysis is remarkably calculation of {\it unidentified} charged hadron FFs up to NNLO. So we need the computation of the SIA cross sections and the time-like DGLAP evolution of the FFs up to NNLO.
To this aim, we use the publicly available {\rm APFEL}~\cite{Bertone:2013vaa} code in which the numerical solution of the time-like evolution equations are performed in the $\overline{\rm MS}$ factorization scheme in $z$-space.
Concerning the zero mass quark assumption, we use zero-mass variable-flavor-number scheme (ZM-VFNS) to account the contributions of heavy flavor.

Some physical parameters are used in the computation of the SIA cross-sections and also in the evolution of FFs. The values of these parameters in our analysis have been chosen as follows: For the heavy flavor masses we use $m_c=1.43$~GeV and $m_b=4.3$~GeV, respectively. We also use $M_Z=91.187$~GeV for the Z-boson mass, and $\alpha_s(M_Z)=0.118$ as a QCD coupling value~\cite{Patrignani:2016xqp}.

In the next section, we briefly highlight the main feature of the {\tt SGK18} FFs analysis, specifically discussing {\tt SGK18} choice of parameterizations of the {\it unidentified} charged hadron FFs at the input scale and the heavy flavor mass scheme. The parameters describing the NLO and NNLO FFs also presented in the next section as well.

\section{ Outline of the {\tt SGK18} FFs analysis } \label{sec:parametrization}

In this section, we present the methodology of {\tt SGK18} FFs analysis, the input functional form and our assumptions we use in this analysis. 
As we mentioned, determination of individual fragmentation functions $D_i^h$ for all quark flavors $i$ as well as gluon into {\it unidentified} charged hadron at NLO and NNLO is the main aim of the present global analysis.
We are also interested in studying the general features of NNLO corrections. As we discussed, the QCD framework for the NNLO corrections are only available for the single electron-positron annihilation among the hard scattering processes and only in the ZM-VFNS. 

We follow the same flexible functional form to parametrize the non-perturbative input FFs at initial scale $Q_0$ used in the series of DSS global QCD analyses~\cite{Anderle:2015lqa,deFlorian:2017lwf,deFlorian:2007aj,deFlorian:2007ekg}. 
In view of this fact, and in order to account the light quark decomposition $q  + \bar{q}$, we assume the following general initial functional form for {\tt SGK18} FFs analysis at a given input scale: 

\begin{eqnarray}\label{parametrization}
D^h_i(z, Q_0^2) = \frac{{\cal N}_i z^{\alpha_i}(1 - z)^{\beta_i}
[1 + \gamma_i(1 - z)^{\delta_i}]}{B[2 + \alpha_i, \beta_i + 1] +
\gamma_i B[2 + \alpha_i, \beta_i +
\delta_i + 1]}, \, \nonumber \\
\end{eqnarray}
%
where $B[a,b]$ is the Euler Beta function which is used to normalize the parameter ${\cal N}_i$.

We should notice here that the standard electron-positron annihilation data sets only provide information on the certain hadron spices summed over the charge, and hence, they are only sensitive to flavor combinations of $q  + \bar{q}$, $i=u + \bar{u}, d + \bar{d}, s+\bar{s}, c+\bar{c}, b+\bar{b}$ and $g$. Since the observables for the {\it unidentified} charged hadron are usually presented for the sum $d \sigma^{h} = d \sigma ^{h^+} + d \sigma^{h^-}$, we only parameterize $D^h$ in our analysis. According to the charge conjugation $D^{h^+}_{q(\bar{q})}  =  D^{h^-}_{\bar{q}(q)}$, we can separate quark and antiquark contributions as

\begin{eqnarray}\label{pelasmines}
D^h_q = D^h_{\bar{q}} = \frac{D^h_{q^+}} {2}.
\end{eqnarray}

Since SIA data is sensitive to the $D_{d+s}$, in {\tt SGK18} FFs analysis, we assume the symmetric fragmentation functions for $d$ and $s$ quark as $D^h_{d + \bar{d}} = D^h_{s + \bar{s}}$.
Moreover, since these data sets are not sensitive to all parameters for the charm and bottom FFs, we assume $\gamma_{c + \bar{c}} = \gamma_{b + \bar{b}} =0$ and $\delta_{c + \bar{c}} = \delta_{b+\bar{b}}=0$. Hence, we choose the most simple functional form for the heavy charm and bottom FFs as follows,

\begin{eqnarray}\label{c and b}
D^h_i(z, Q_0^2) = \frac{{\cal N}_i
z^{\alpha_i}(1 - z)^{\beta_i}} {B[2 + \alpha_i,
\beta_i + 1]}, ~~~i = c + \bar{c}, b + \bar{b} \,.
\end{eqnarray}

The parameter $\gamma_g$ for the gluon FFs is basically unconstrained by the analyzed data sets, and in order to get the best fit, we decided to keep it fixed at $\gamma_g = 70$ for both {\tt SGK18} NLO and NNLO analyses. We discuss in section~\ref{sec:Results} that the gluon FF obtained in {\tt SGK18} analysis is slightly different from the {\tt DSS07} analysis which used the SIDIS and hadron collider data. The proton-antiproton data from {\tt CDF}~\cite{Aaltonen:2009ne,Abe:1988yu} experiment at SLAC, the proton-proton data from {\tt CMS}~\cite{Chatrchyan:2011av,CMS:2012aa} and {\tt ALICE}~\cite{Abelev:2013ala} experiments at CERN carry a large amount of information on the gluon FF and could constrain it well enough. However, the data from single-inclusive charged hadron production in $e^-e^+$ annihilation is the major source of exprimental data in our analysis.

We should mention here that in {\tt SGK18} FFs analyses, the initial scale for input parametrization is $Q_0=5~$GeV for all parton species.
Since the value of bottom mass in our analysis is $m_b = 4.3$~GeV, this initial scale is above bottom threshold. In addition, this value for $Q_0$ is below the lowest center-of-mass energy of analyzed data sets, $s = \sqrt{14}$~GeV. 
Since time-like matching conditions are unknown at NNLO, with this value for $Q_0$, it is not require heavy quark threshold as well as the matching in the evolution between the initial scale and the data scale. Therefore, in our analysis the number of active flavor keep fixed to the $n_f=5$.

%
\section{$\chi^2$ minimization and calculation method of errors } \label{sec:errorcalculation}
%

The parameters describing the {\it unidentified} charge hadron FFs presented in Eqs.~\eqref{parametrization} and \eqref{c and b} are determined using a standard $\chi^2$ minimization method. 
The total $\chi^2$ is calculated in comparison with the single-inclusive charged hadron production data sets in electron-positron annihilation for the {\it unidentified} charge hadron FFs.
In order to calculate the $\chi^2$, the theoretical predictions should be obtained at the same experimental $z$ and $\mu^2=Q^2$ points.
As we mentioned, the $\mu^2=Q^2$ evolution is calculated by the well-known DGLAP evolution equations.

In order to calculate the total $\chi^2 (\{\eta_i\})$ for independent sets of fit parameters $\{\eta_i\}$, one can use the following standard $\chi^2$ definition: 

\begin{eqnarray}\label{eq-chi2}
\chi^2 (\{\eta_i\}) = \sum_{i}^{ n^{data} }
\big( \frac{ {\cal E}_i -
{\cal T}_i (\{\eta_i\}) } { \delta {\cal E}_i } \big )^2 \,,
\end{eqnarray}
%
where ${\cal E}_i$ is the measured value of a given observable and ${\cal T}_i$ is the corresponding theoretical estimate for a given set of parameters $\{\eta_i\}$ at the same experimental $z$ and $\mu^2 = Q^2$ points.
The experimental errors associated with this measurements are calculated from systematic and statistical errors added in quadrature, $(\delta E_i)^2 = (\delta E^{sys}_i)^2 + (\delta E^{stat}_i)^2$. The optimization is done by the CERN program MINUIT~\cite{James:1994vla}.

Since most single-inclusive charged hadron production data in $e^- e^+$ annihilation come with additional information on the fully correlated normalization uncertainty, the above simple $\chi^2$ definition need to be modified in order to account for such normalization uncertainties.
Hence, the modified function is given by,

\begin{eqnarray}\label{eq-chi2global}
\chi_{global}^2 (\{\eta_i\}) & = &
\sum_{n=1}^{n^{exp}} \left( \frac{1 -{\cal N}_n }{\Delta{\cal N}_n}\right)^2 + \nonumber \\ 
&&\sum_{j=1}^{N_n^{data}} \left(\frac{ ( {\cal N}_n
\, {\cal E}_j^{data} - {\cal T}_j^{theory}(\{\eta_i\})}
{{\cal N}_n \, \delta {\cal E}_j^{data}} \right)^2 \,, \nonumber \\ 
\end{eqnarray}
%
where $n^{\rm exp}$ corresponds to the individual experimental data sets for the $n^{th}$ experiment, and $N^{\rm data}_n$ refers to the number of data points in each data set.
The normalization factors $\Delta {\cal N}_n$ in above equation can be fitted along with the fitted parameters $(\{\eta_i\})$ of Eqs.~\eqref{parametrization} and \eqref{c and b}  and then keep fixed.
In order to illustrate the effects arising from the use of the different single-inclusive charged hadron production data sets, in Table.~\ref{tab:datasets}, we have shown the obtained $\chi/n^{data}$ for each data sets at NLO and NNLO accuracy.
This table illustrates the quality of {\tt SGK18} NLO and NNLO QCD fits to single-inclusive charged hadron production data in terms
of the individual $\chi^2$-values obtained for each experiment.
The total $\chi^2/N_{pts}$ for the {\tt SGK18} fits can be found in this table as well.
We obtained  1.64 and 1.62 for our NLO and NNLO analyses, respectively. 

This section also focuses on the uncertainties of the parameters in Eqs.~\eqref{parametrization} and \eqref{c and b} to judge the quality of {\tt SGK18} QCD fits.
In order to determine the uncertainties of {\it unidentified} charged hadron FFs as well as the corresponding observable, we apply the ``Hessian'' method by choosing a particular value of $\Delta \chi^2 =1$.
This will provide a clear and comprehensive picture of the uncertainty characteristic of resulting FFs.

The determination of the size of uncertainties using the ``Hessian'' method is based on the correspondence between the confidence level (C.L.) ${\cal P}$ and $\chi^2$ with the number of fitting parameters $N$.
The C.L. is given by,

\begin{equation}\label{CL}
{\cal P} = \int_0^{\Delta \chi^2}\frac{1}{2\, \Gamma(N/2)}
\left( \frac{\zeta^2}{2} \right)^{\frac{N}{2} - 1}
e^{\left(-\frac{\zeta^2} {2} \right)} d\, \zeta^2 \,,
\end{equation}
%
where $\Gamma$ is the Gamma function.
The value of $\Delta \chi^2$ in Eq.~\eqref{CL} is taken so that the C.L. becomes the one-$\sigma$-error range, namely $P=0.68$. The value for the $\Delta \chi^2$ is then numerically calculated by using this equation.

Having at hand the value for $\Delta \chi^2$ and the derivatives of given observables with respect to the fitted parameters \{$\eta_i$\} ($i$=1, 2, ..., $N$), the Hessian approach provides the uncertainties of desired observables ${\cal O}$ as,

\begin{equation}\label{Uncer}
[\Delta {\cal O}_i]^2  =  \Delta \chi^2 \sum_{m,n}
\left( \frac{\partial {\cal O}_i (\eta)}
{\partial \eta_m}  \right)_{\hat \eta}
C_{m,n}	\left( \frac{\partial {\cal O}_i (\eta)}
{\partial \eta_n}  \right)_{\hat \eta} \,,
\end{equation}
%
where $C_{m,n}$ is the inverse of the Hessian matrix which can be obtained by running the CERN program library MINUIT~\cite{James:1994vla}.

For estimation of uncertainties at an arbitrary $Q^2$ which is an attributive function of the input parameters,
the obtained gradient terms are evolved by the well-known DGLAP evolution kernel.
In next section we show that the {\tt SGK18} FFs uncertainties determination as well as the fitting methodology can correctly propagate the experimental uncertainty of the single-inclusive charged hadron production data into the uncertainties of the {\tt SGK18} FFs.

%
\section{ ${\tt SGK18}$ fit results} \label{sec:Results}
%

In this section, we present the {\tt SGK18} numerical results for the {\it unidentified} charged hadron $D^h$ FFs obtained from the global analysis of SIA data.
Firstly, we present the parameters of the optimum QCD fits describing the {\it unidentified} charged hadron and then we present the {\tt SGK18} FFs results for different partons at NLO and NNLO accuracy in pQCD.
Then, {\tt SGK18} results for FFs are compared to the {\tt DSS07} FFs for {\it unidentified} charged hadron.
Secondly, the uncertainty bands at NLO and NNLO accuracy are compared and the improvement of the FFs calculations due to the inclusion of NNLO QCD corrections are discussed.
Next, {\tt SGK18} theoretical predictions for the total cross sections and all different tagged cross sections are compared with the analyzed SIA experimental data sets.
Finally, we present our theoretical uncertainties from the variation of the renormalization and factorization scales.

%
\subsection{ {\tt SGM18} FFs and comparison with DSS FF sets } 
%

Even though we mainly interested in a precise extraction of  {\it unidentified} charged hadron FFs  $D^h$ at NNLO accuracy, we also present the results of our analysis at NLO approximation.
As we will discuss in this section, the significantly better NNLO uncertainty highlights the importance of higher order correction in our QCD analysis.  
In addition our NLO results can be used in calculation of observable which are limited to the NLO corrections.

The 21 best fit parameters describing the optimum NLO and NNLO {\it unidentified} charged hadron $D^h$ FFs are given in Tables~\ref{tab:NLO} and \ref{tab:NNLO}.

%
\begin{table*}[t]
	\caption{\label{tab:NLO} Fit parameters for the fragmentation of quarks and gluon into the $D^h$-meson at NLO accuracy.
		The starting scale is taken to be $Q_{0}=5$~GeV for all parton species.
		The values labeled by (*) have been fixed after the first minimization, since the available SIA data dose not constrain all unknown fit parameters well enough.  }
	\begin{ruledtabular}
		\begin{tabular}{cccccc}
			flavor $i$ & ${\cal N}_i$ & $\alpha_i$ & $\beta_i$ & $\gamma_i$ & $\delta_i$                             \\ \hline
			$u+ \overline{u}$ & $0.332 \pm 0.006$ & $-0.539 \pm 0.171$ & $1.499 \pm 0.114$ & $4.882 \pm 1.891$ & $4.082 \pm 0.703$        \\
			$d+ \overline{d}$ & $0.411\pm 0.007$ & $-0.829 \pm 0.084$ & $2.622 \pm 0.361$ & $1.750 \pm 1.388$ & $2.411 \pm 0.721$   \\
			$g$ & $0.259 \pm 0.009$ & $0.256 \pm 0.066$ & $1.165 \pm 0.144$ & $70^*$ & $10.847 \pm 0.597$   \\
			$c+ \overline{c}$ & $0.191 \pm 0.003$ & $-0.845 \pm 0.031$ & $4.369 \pm 0.141$ & $0.0$ & $0.0$        \\
			$b+ \overline{b}$ & $0.149 \pm 0.002$ & $ -0.626 \pm 0.025$ & $7.291 \pm 0.170$ & $0.0$ & $0.0$   \
		\end{tabular}
	\end{ruledtabular}
\end{table*}

%
\begin{table*}[t]
	\caption{\label{tab:NNLO} Same as Table~\ref{tab:NLO} but for the NNLO analysis.   }
	\begin{ruledtabular}
		\begin{tabular}{cccccc}
			flavor $i$ & ${\cal N}_i$ & $\alpha_i$ & $\beta_i$ & $\gamma_i$ & $\delta_i$                             \\ \hline
			$u+ \overline{u}$ & $0.312 \pm 0.004$ & $0.158 \pm 0.034$ & $1.659 \pm 0.045$ & $22.609 \pm 1.560$ & $6.347 \pm 0.152$        \\
			$d+ \overline{d}$ & $0.476 \pm 0.003$ & $-1.583 \pm 0.013$ & $3.037 \pm 0.040$ & $-0.919 \pm 0.005$ & $1.364 \pm 0.125$   \\
			$g$ & $0.146 \pm 0.002$ & $ 0.328 \pm 0.0319$ & $9.410 \pm 0.162$ & $70^*$ & $1.072 \pm 0.165$   \\
			$c+ \overline{c}$ & $0.208 \pm 0.002$ & $-0.93 \pm 0.0193$ & $3.746 \pm 0.076$ & $0.0$ & $0.0$        \\
			$b+ \overline{b}$ & $0.169 \pm 0.001$ & $ -0.833 \pm 0.012$ & $5.584 \pm 0.057$ & $0.0$ & $0.0$   \
		\end{tabular}
	\end{ruledtabular}
\end{table*}

\begin{figure*}[htb]
	\begin{center}
		\vspace{0.50cm}
		\resizebox{0.48\textwidth}{!}{\includegraphics{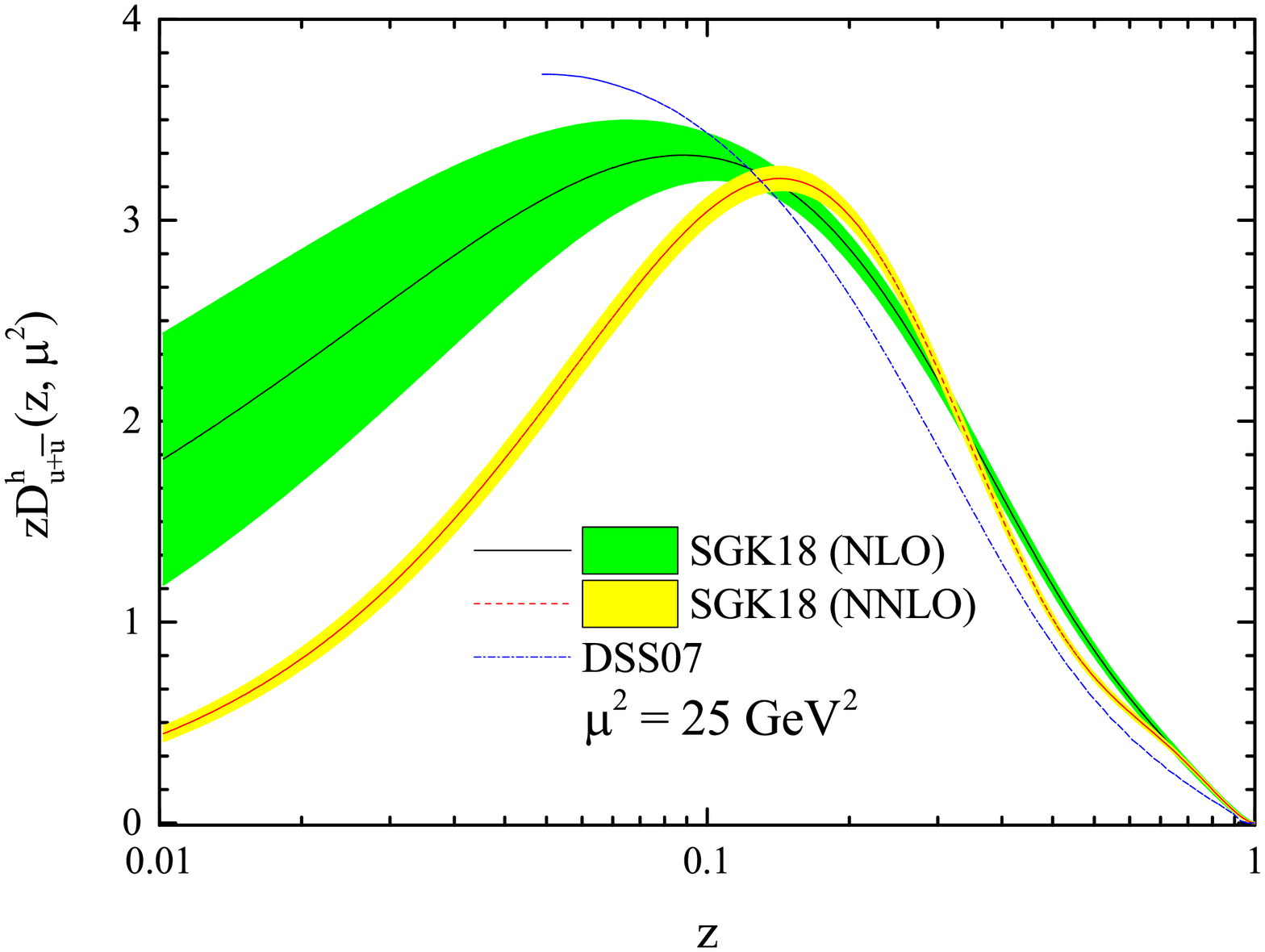}}  
		\resizebox{0.48\textwidth}{!}{\includegraphics{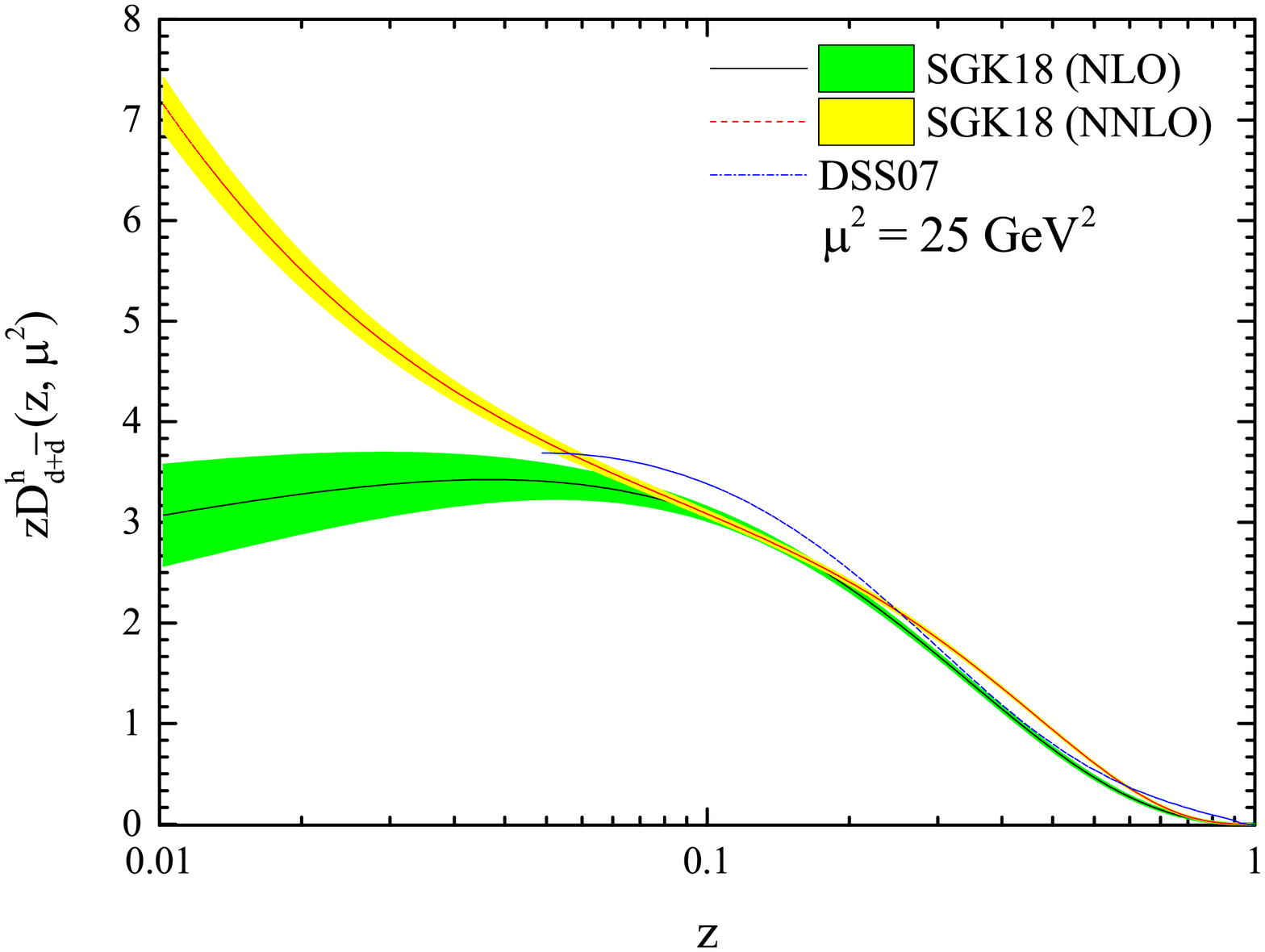}} 
		\resizebox{0.48\textwidth}{!}{\includegraphics{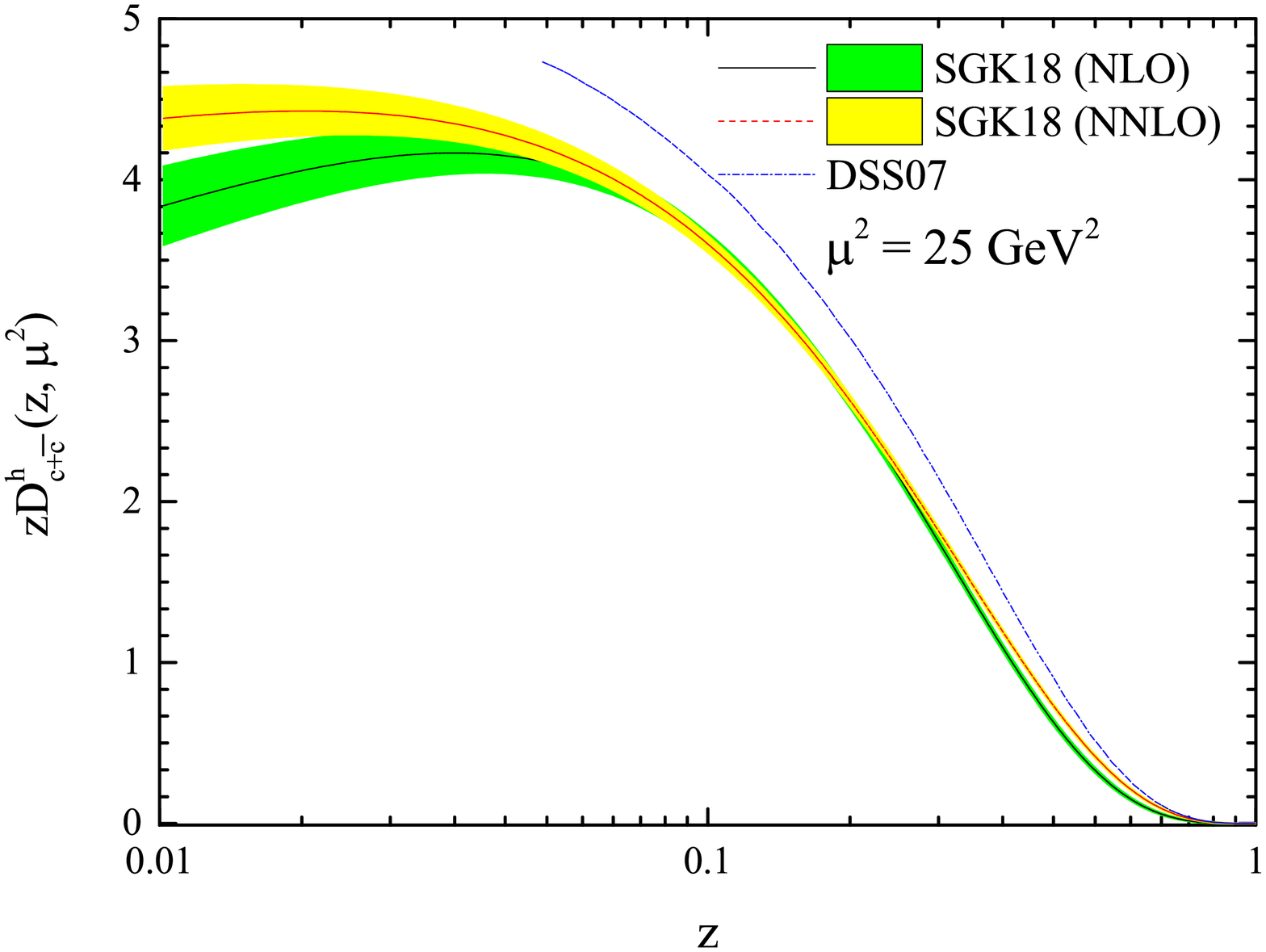}}  
		\resizebox{0.48\textwidth}{!}{\includegraphics{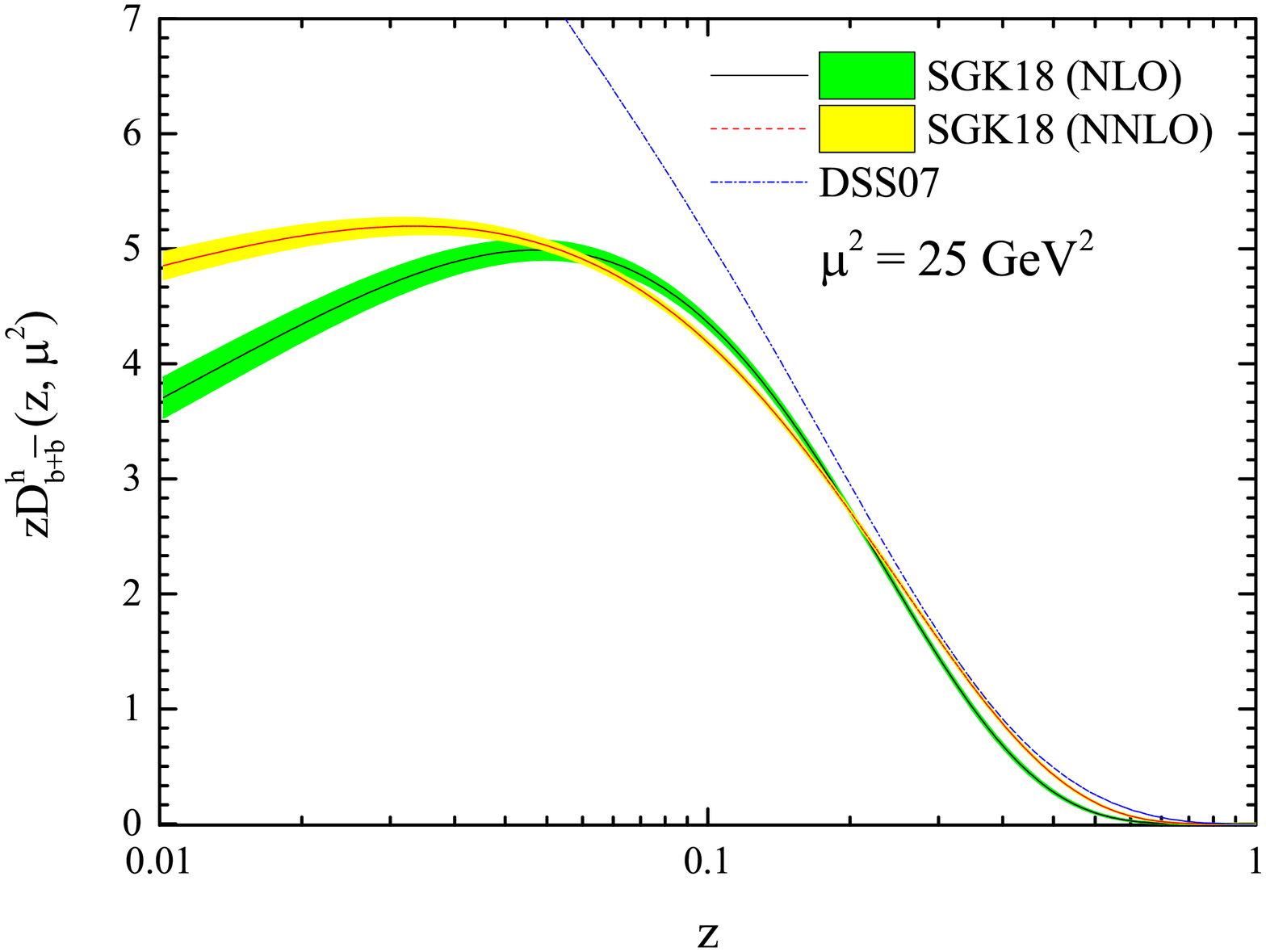}} 
		\resizebox{0.48\textwidth}{!}{\includegraphics{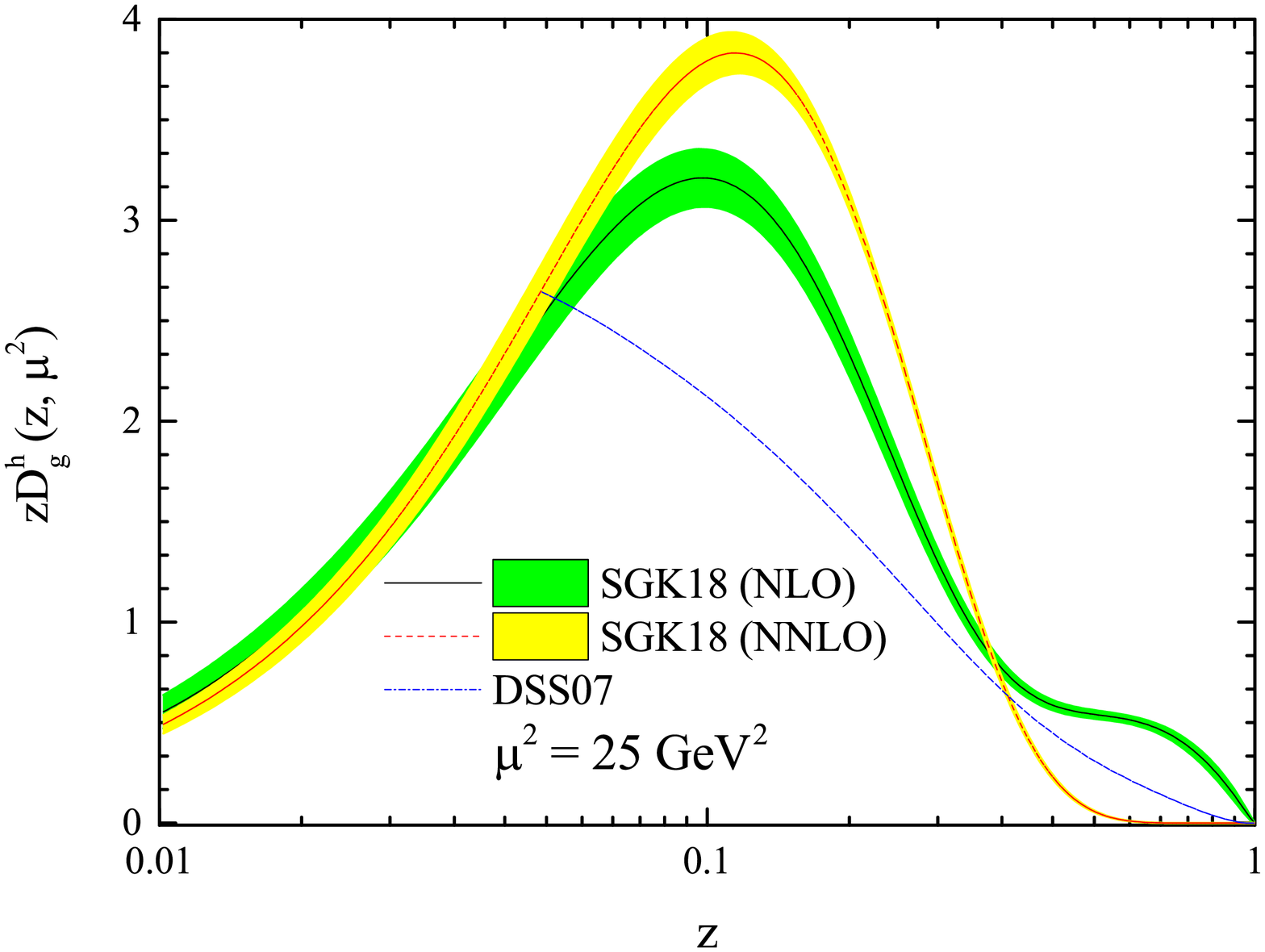}} 
		\caption{ {\tt SGK18} fragmentation densities and their uncertainties (shaded bands) for $zD^h_i$ at the initial scale of $Q_0^2=25$~GeV$^2$ for $u + \bar{u}$, $d + \bar{d}$, $c + \bar{c}$, $b + \bar{b}$ and $g$ both at NLO (solid lines) and NNLO (dashed lines). Our results have also been compared with the {\tt DSS07} (dot-dashed lines) results at NLO~\cite{deFlorian:2007ekg}.} \label{fig:FFsQ0}
	\end{center}
\end{figure*}

\begin{figure*}[htb]
	\begin{center}
		\vspace{0.50cm}
		\resizebox{0.48\textwidth}{!}{\includegraphics{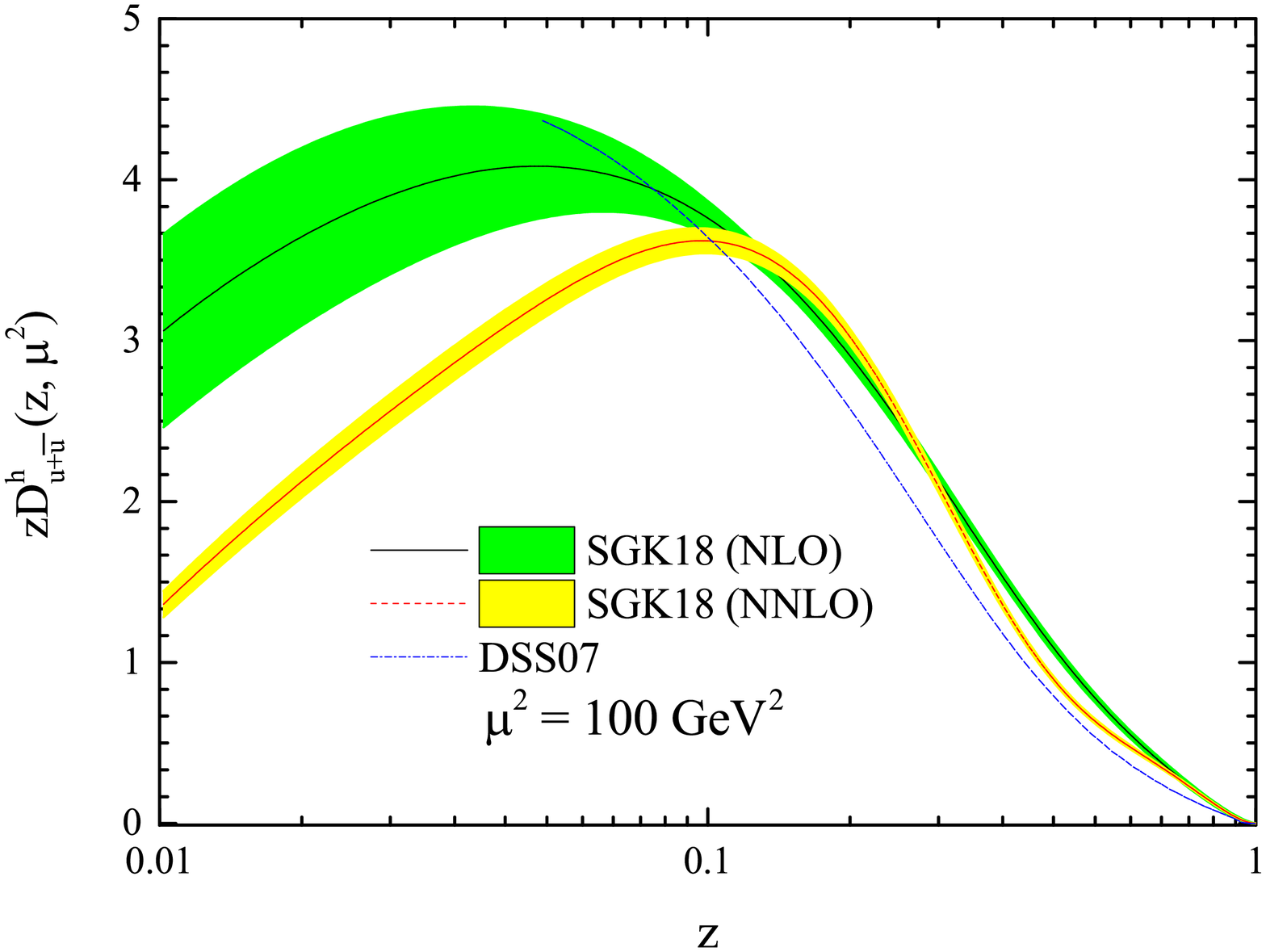}}  
		\resizebox{0.48\textwidth}{!}{\includegraphics{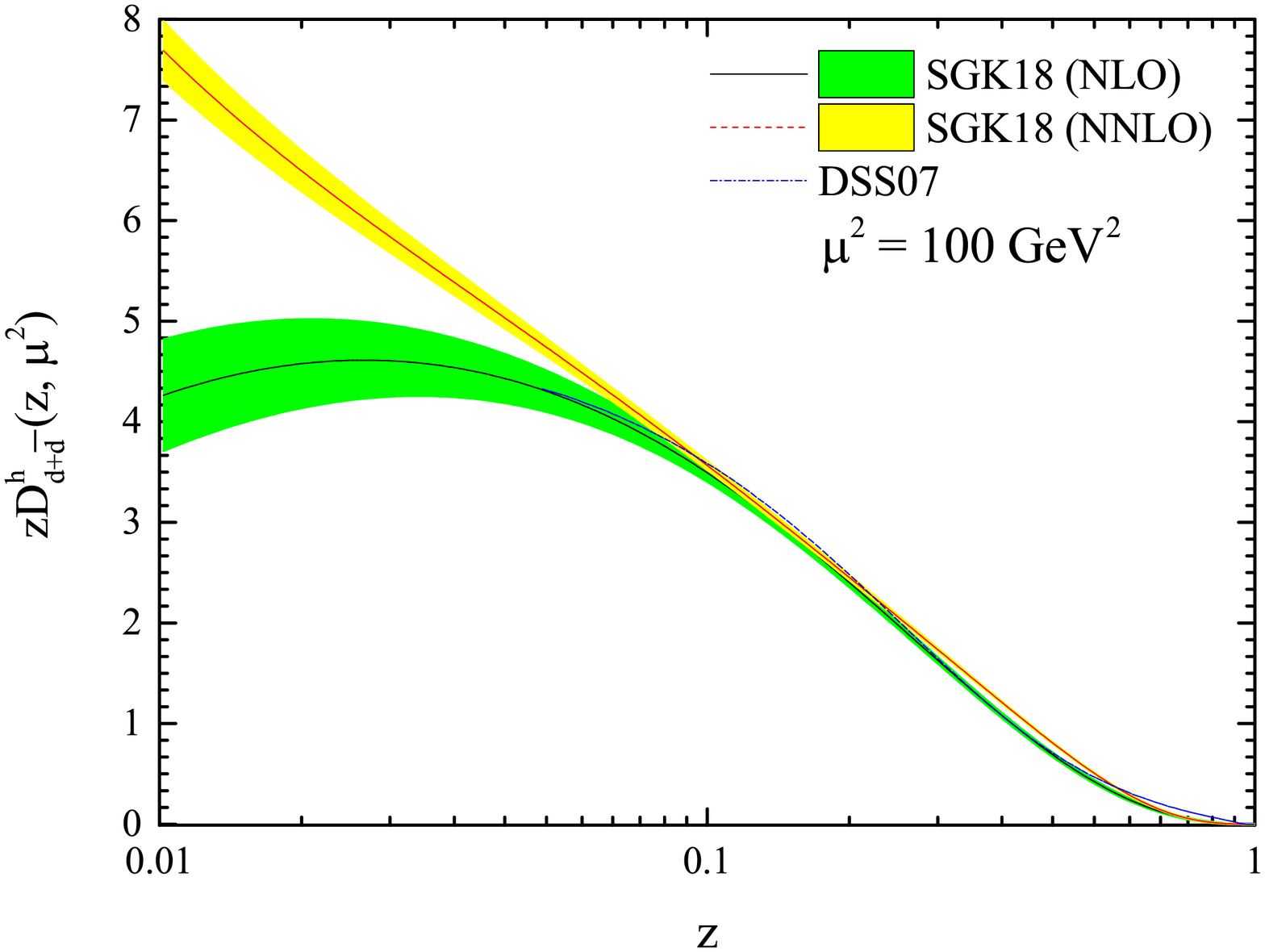}} 
		\resizebox{0.48\textwidth}{!}{\includegraphics{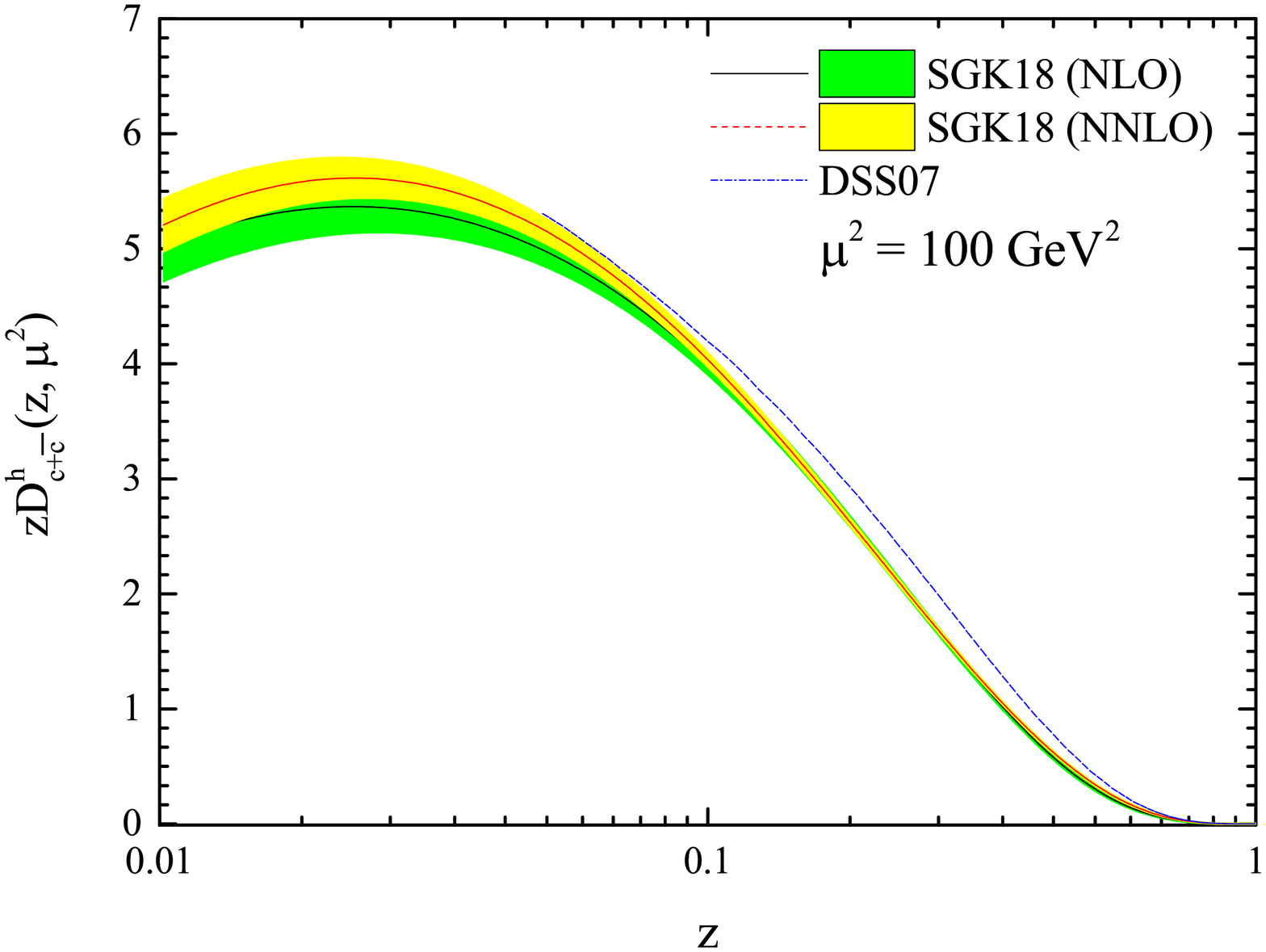}}  
		\resizebox{0.48\textwidth}{!}{\includegraphics{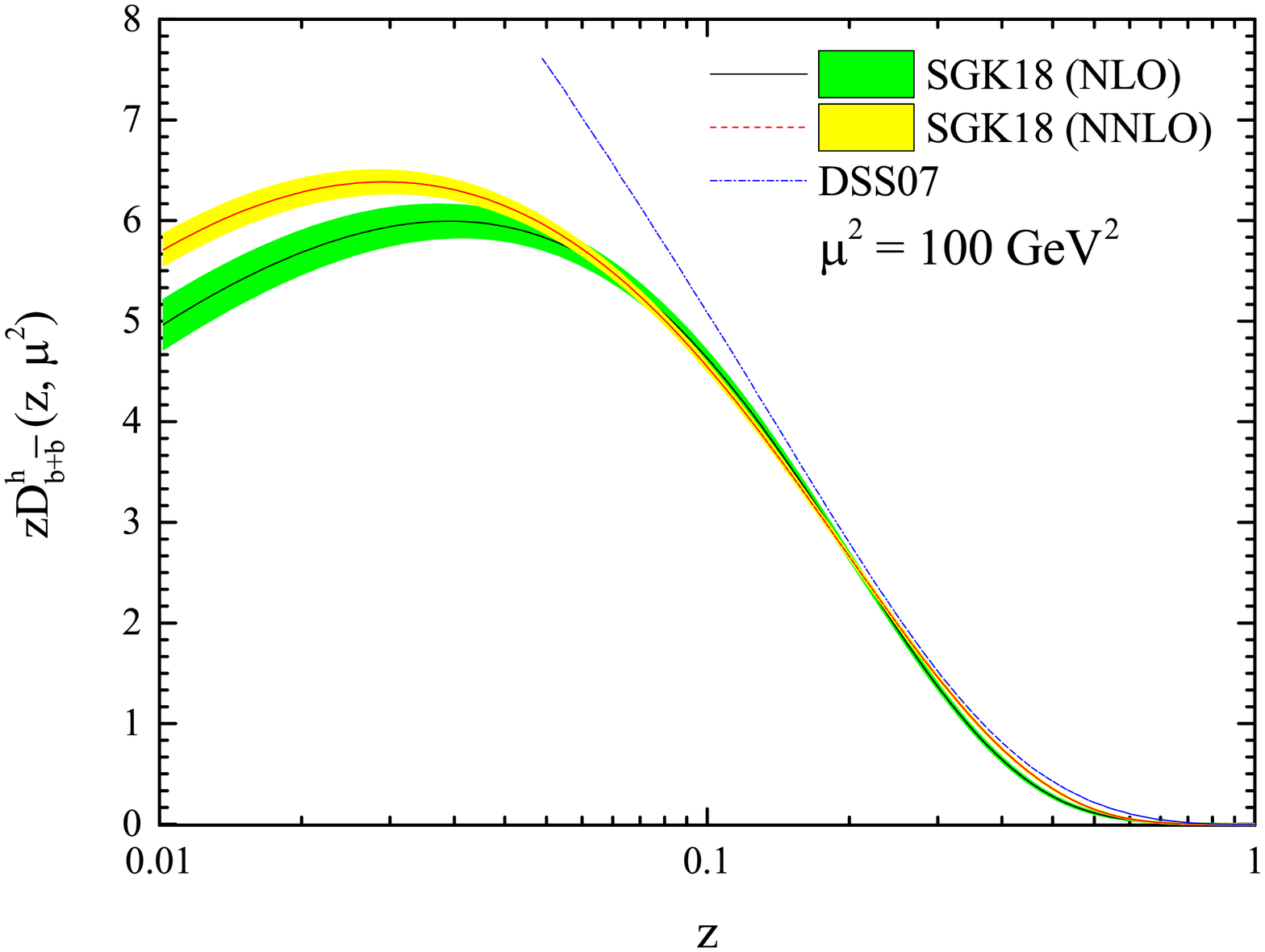}} 
		\resizebox{0.48\textwidth}{!}{\includegraphics{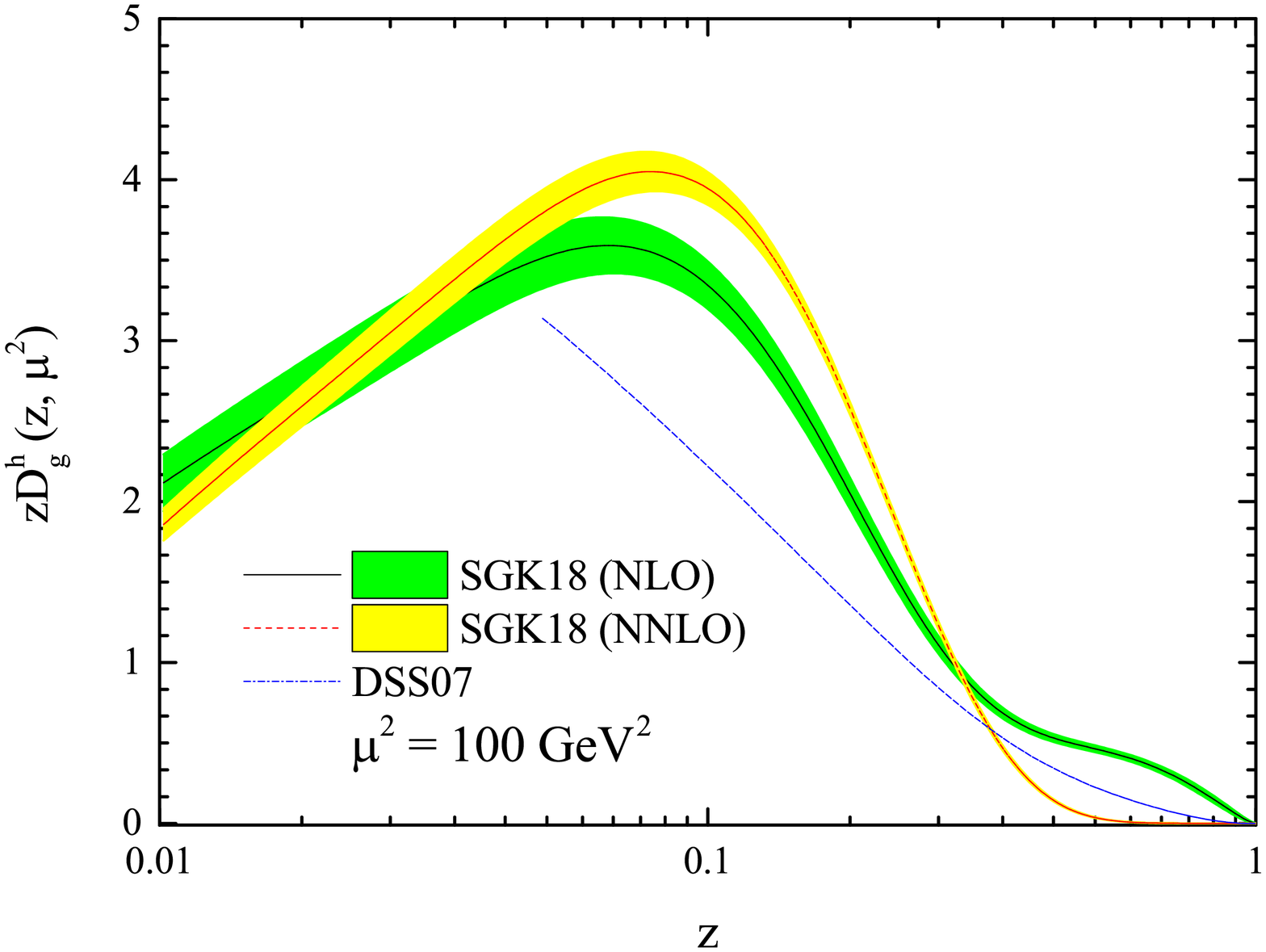}} 
		\caption{ Same as Fig.~\ref{fig:FFsQ0} but for $Q^2=100$~GeV$^{2}$. }
		\label{fig:FFs100}
	\end{center}
\end{figure*}

\begin{figure*}[htb]
	\begin{center}
		\vspace{0.50cm}
		\resizebox{0.48\textwidth}{!}{\includegraphics{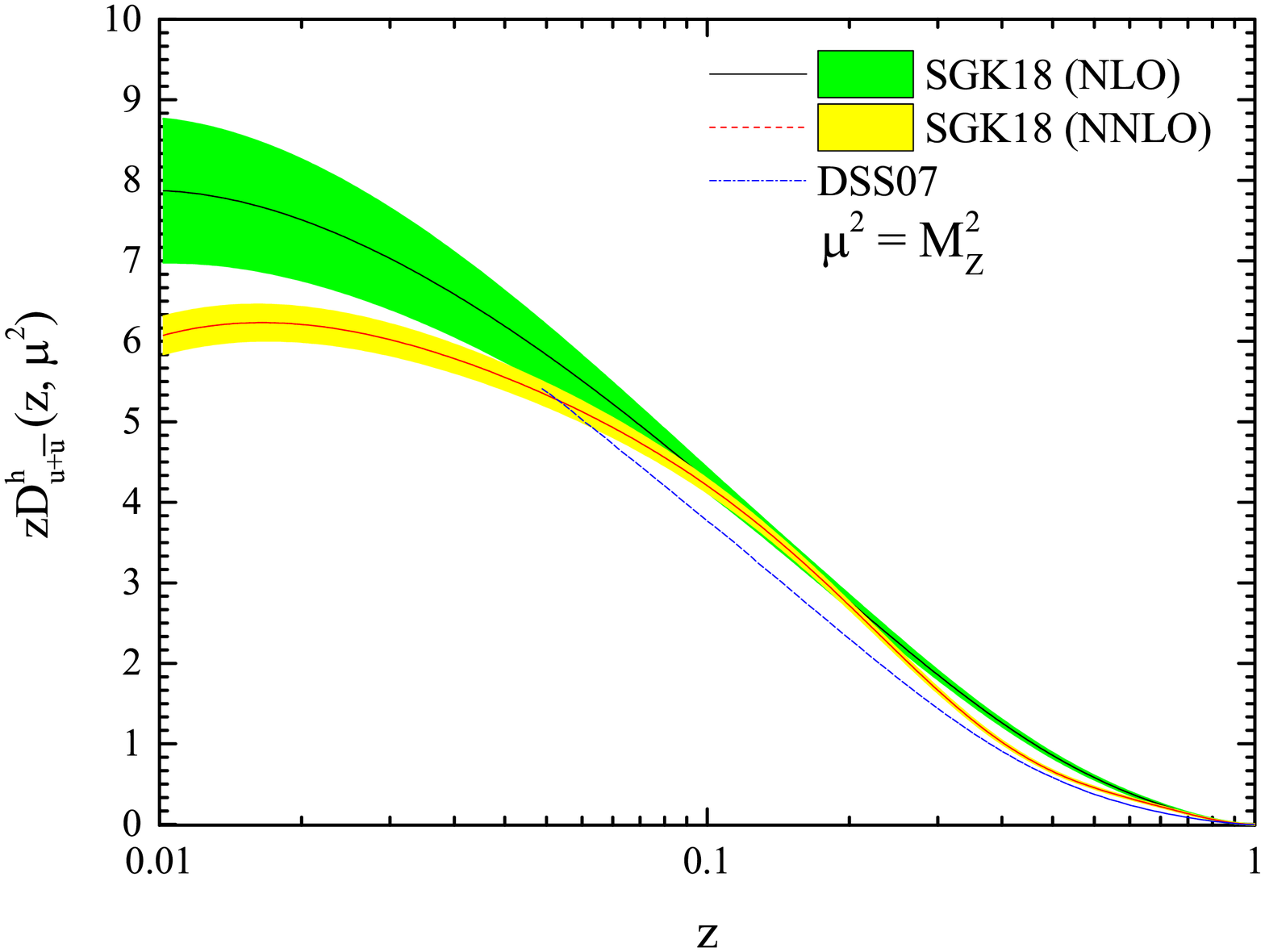}}  
		\resizebox{0.48\textwidth}{!}{\includegraphics{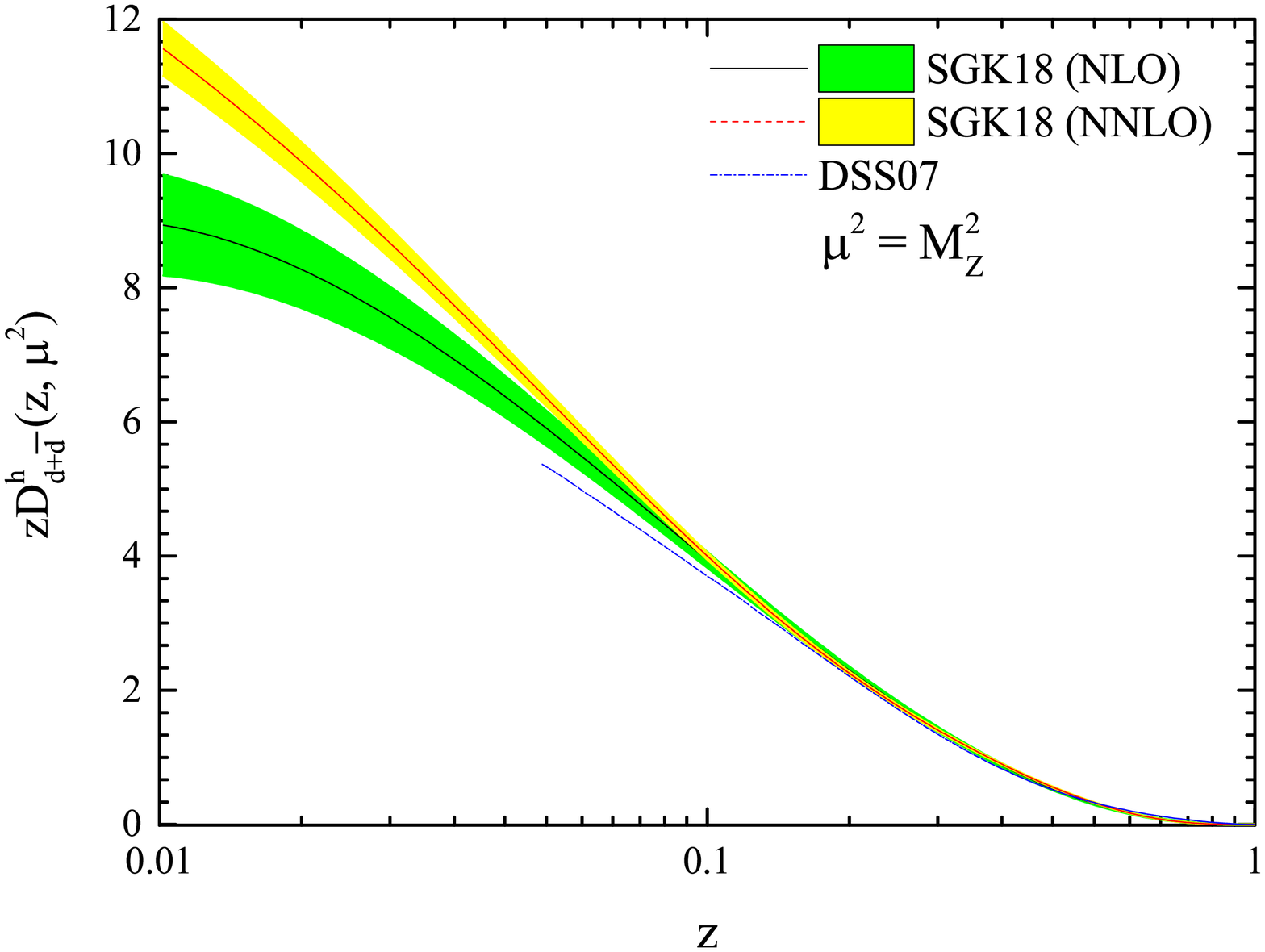}} 
		\resizebox{0.48\textwidth}{!}{\includegraphics{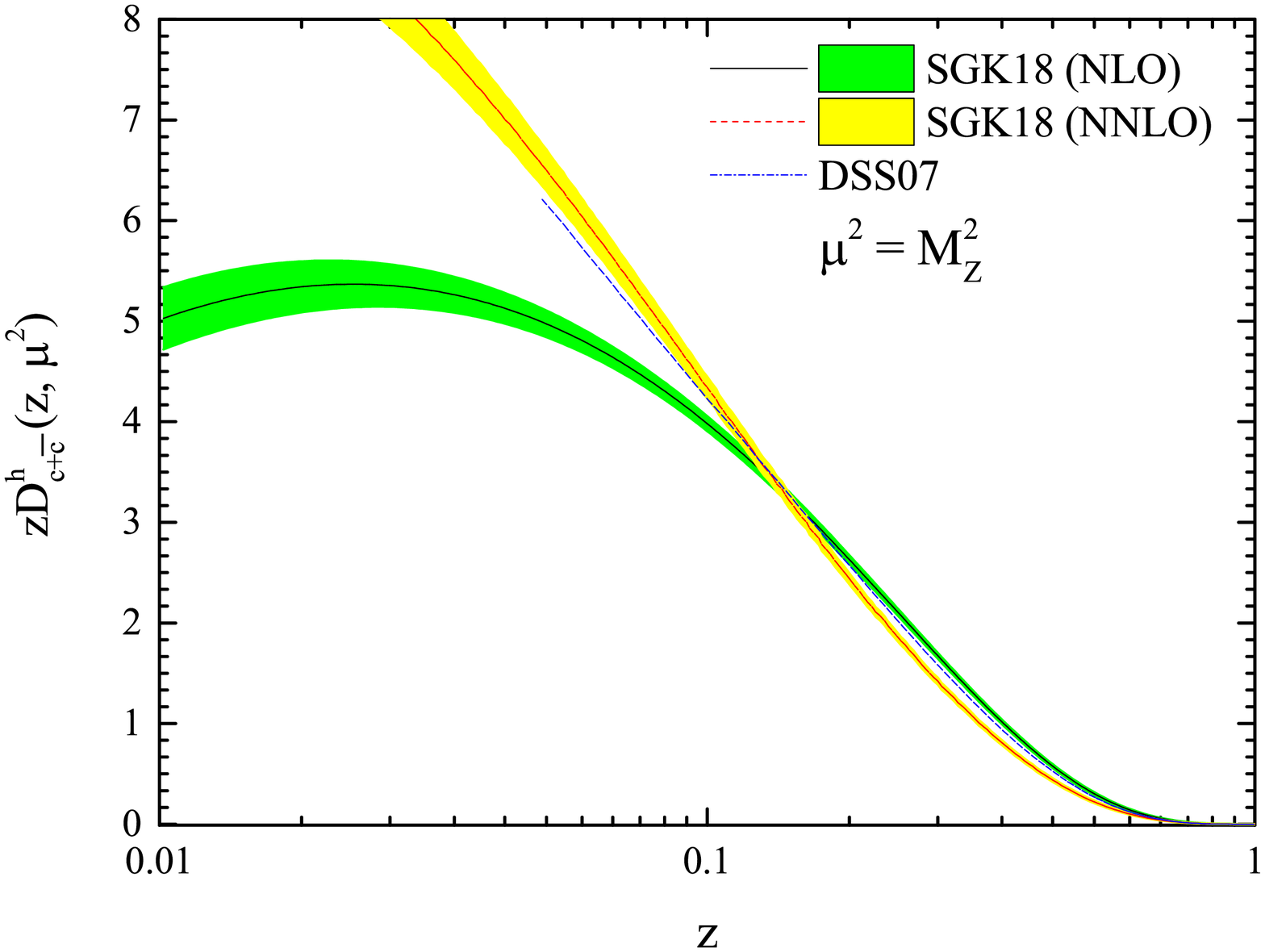}}  
		\resizebox{0.48\textwidth}{!}{\includegraphics{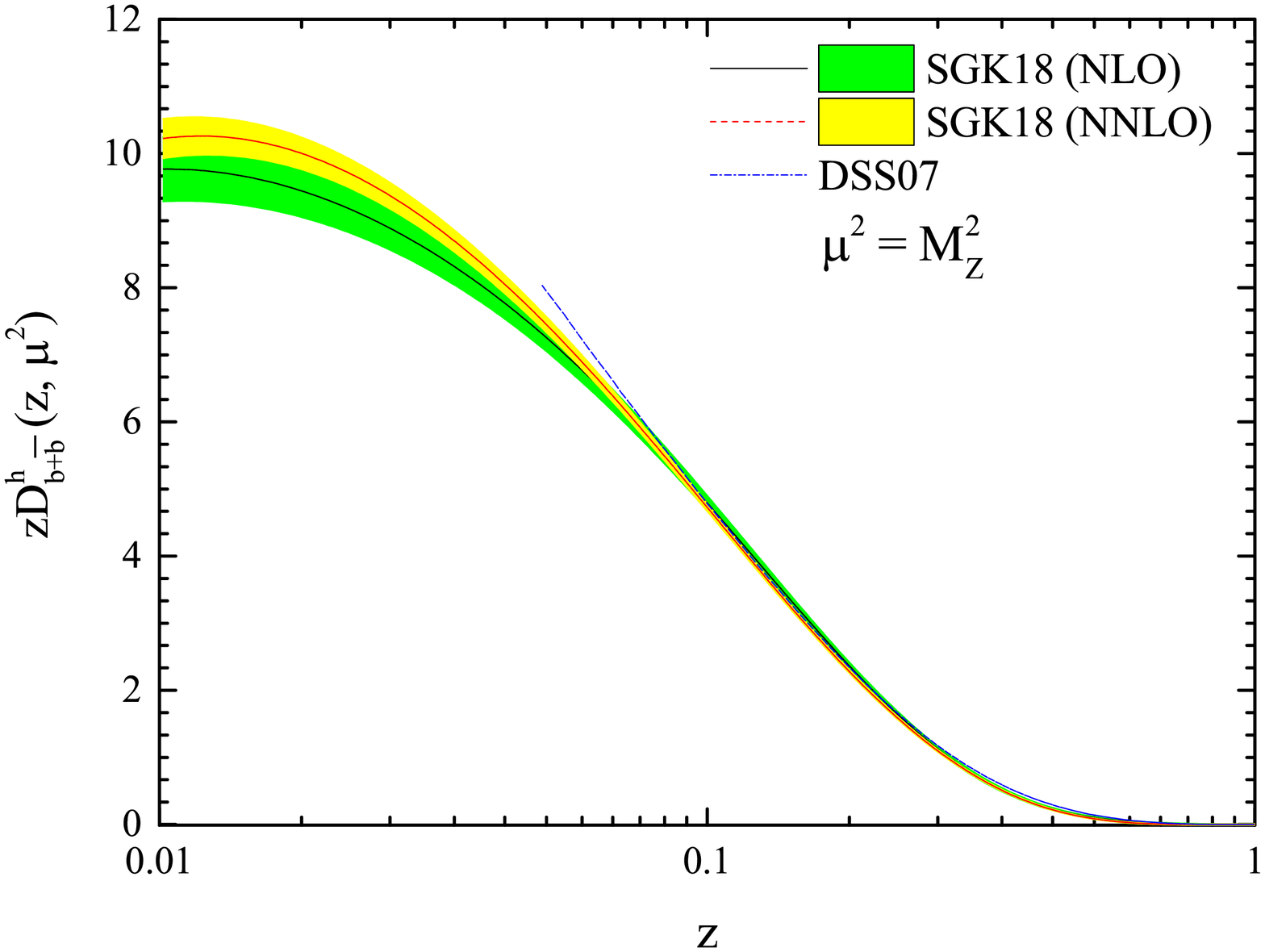}} 
		\resizebox{0.48\textwidth}{!}{\includegraphics{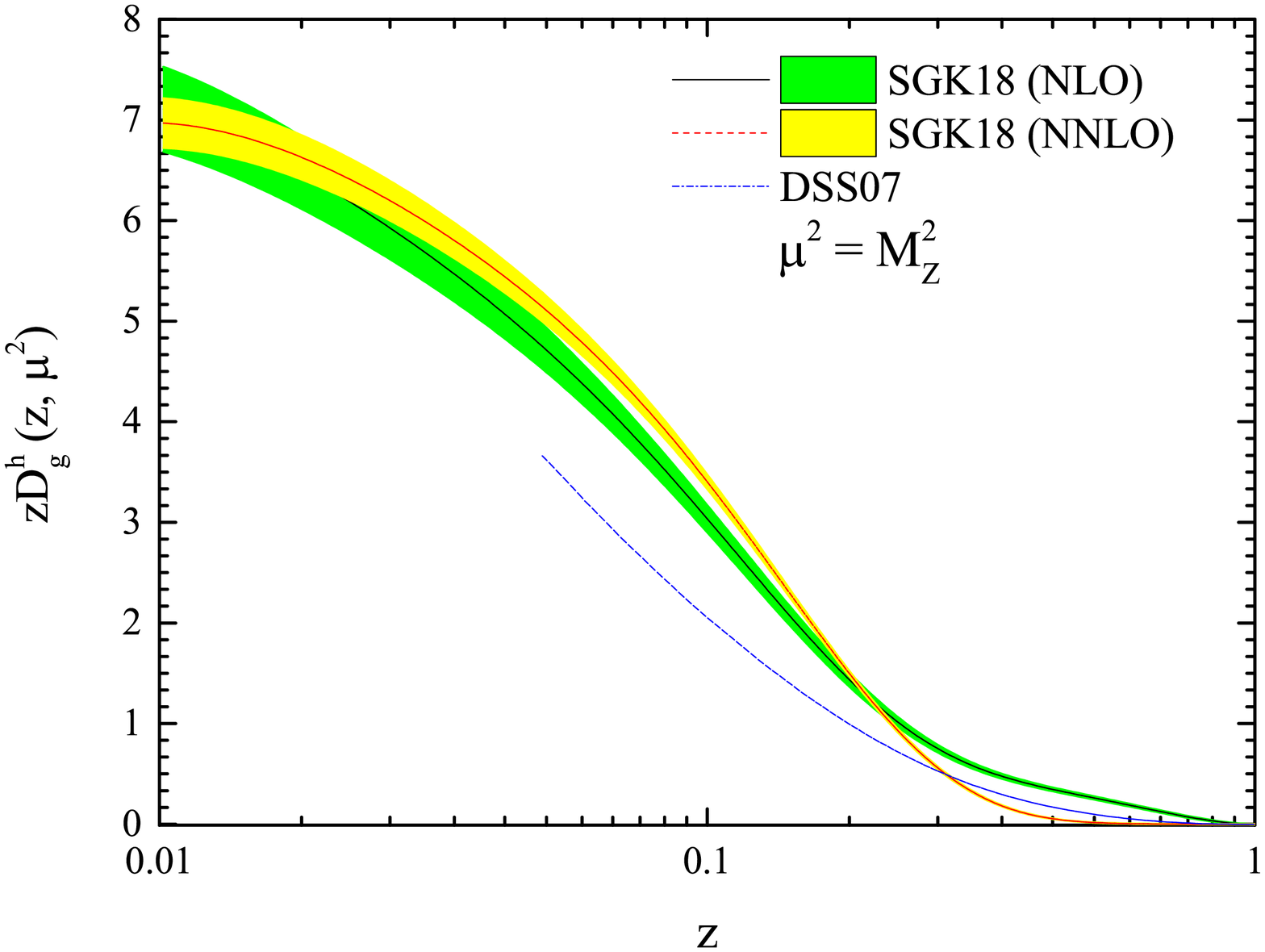}} 
		\caption{ Same as Fig.~\ref{fig:FFsQ0} but for $Q^2=M_Z^2$.  } \label{fig:FFsMZ}
	\end{center}
\end{figure*}

The {\tt SGK18} $D^h$ fragmentation functions and their uncertainties have been presented in Fig.~\ref{fig:FFsQ0} at NLO (solid lines) and NNLO (dashed lines) accuracy for $Q^2$ = 25 GeV$^2$. The resulting NLO and NNLO {\tt SGK18} $D^h$ fragmentation functions and their uncertainties evolved to the scale of 100 GeV$^2$ and $M_Z^2$ have also been illustrated in Figs.\ref{fig:FFs100} and \ref{fig:FFsMZ}. We have included the one-$\sigma$ uncertainty bands in our analysis.


In these figures, we have also compared {\tt SGK18} FFs to the central value of {\tt DSS07} FFs analysis~\cite{deFlorian:2007ekg} (dot-dashed lines) at NLO. The uncertainty bands of the {\tt DSS07} results are not shown in our analysis because they are not available.  Recently, a preliminary determination of the {\it unidentified} charged hadron FFs has been reported in Ref.~\cite{Nocera:2017gbk} at NLO but the grid files are not available. The impact of higher order QCD corrections on the reduction of FFs uncertainties at NNLO accuracy in comparison with the NLO analysis can be seen from these figures. 
As one can see from Figs.~\ref{fig:FFsQ0}, \ref{fig:FFs100} and \ref{fig:FFsMZ}, the uncertainty bands for all quark flavors as well as gluon decrease remarkably at NNLO which indicates that the higher order perturbation QCD corrections increase the precision of the calculations.

Let's discuss the results in more details. Focusing on Fig.~\ref{fig:FFsQ0}, one can clearly see that the results of {\tt SGK18} for light and heavy flavor FFs are similar to the ones obtained from {\tt DSS07} analysis, especially at larger values of the momentum fraction $z$.
However, some noticeable differences can be seen in the gluon FFs. As one can see, the contribution of gluon FF in our analysis is significantly larger than the {\tt DSS07} one.
A main reason for this difference is base on the fact that in {\tt SGK18} analysis only SIA data 
have been included, while the {\tt DSS07} included both the electron-positron SIA and proton-proton collision data in their analysis.
Since the collider data could directly effect the determination of gluon FF, this clear difference observed between our gluon FF result and {\tt DSS07} one is expectable. 
In addition, we should mentioned  this fact that the initial scales used in these two models are different. The {\tt SGK18} initial scale has been chosen to be $Q_0^2=25$~GeV$^2$, while the {\tt DSS07} initial scale is $Q_0^2=1$~GeV$^2$ for light quarks and gluon, and $Q_0^2=m_c^2$ for charm quark FF as well as $Q_0^2=m_b^2$ for the bottom quark FF. 

In addition to the points mentioned above, we have excluded the experimental data below $z_{min}=0.02$ for data at center-of-mass energy of $\sqrt{s}=M_Z$ and $z_{min}=0.075$ for $\sqrt{s}<M_Z$, while the {\tt DSS07} excluded the data points below $z=0.1$ for all analyzed data sets.
Consequently, the number of SIA data points in {\tt DSS07} is 236, but in our analysis there are 474  data points. Because of this difference for the kinematic cuts at small $z$, the most discrepancy is seen at $z<0.1$. In general, the differences become larger towards smaller values of $z$, $z \to 0.01$, which has been already observed for all parton species.
 
The {\tt SGK18} FFs at higher values of $Q^2=100$~GeV$^2$ and $Q^2=M_Z^2$, have been displayed in Figs.~\ref{fig:FFs100} and \ref{fig:FFsMZ}, respectively. It can be conclude from these two figures that as the scale of energy increases, the difference between the {\tt SGK18} gluon FF and {\tt DSS07} is decreased. Furthermore, the behavior of our light and heavy FFs and the {\tt DSS07} ones are slightly in good agreement.

Let us turn to the discussion on the resulting FFs and their uncertainties by focusing on the inclusion of higher order QCD corrections. According to the {\tt SGK18} global fit of SIA data, there are some noticeable features that improve global fit at NNLO in comparison to the NLO.
First, as one can see from Table.~\ref{tab:datasets}, the improvement of the $\chi ^2/{\rm d.o.f}$ from NLO to NNLO is slightly better. Actually, in our analysis the value of $\chi ^2/{\rm d.o.f}$ reduces from $1.64$ at NLO to $1.62$ at NNLO approximation.
Second, the size of the {\tt SGK18} FFs uncertainties remarkably decrease at NNLO in comparison to our NLO analysis. One can see the reduction of uncertainties for all determined quark flavors as well as gluon at all three scales of energy shown in Figs.~\ref{fig:FFsQ0}, \ref{fig:FFs100} and \ref{fig:FFsMZ}. One can conclude that the effects arising due to the inclusion of higher order QCD corrections significantly decrease the obtained error bands.

\begin{figure*}[htb]
	\begin{center}
		\vspace{0.50cm}
		\resizebox{0.48\textwidth}{!}{\includegraphics{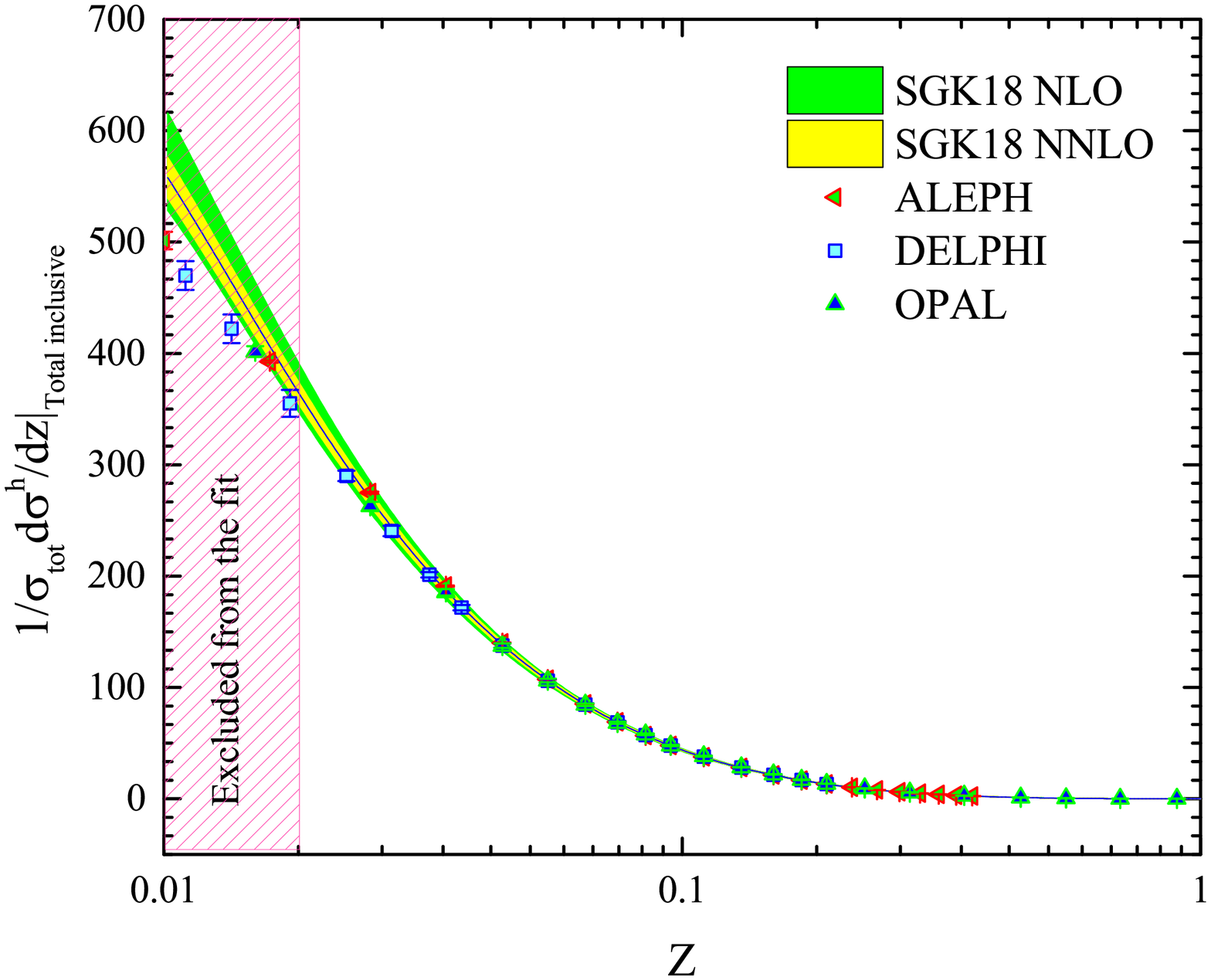}}  
		\resizebox{0.48\textwidth}{!}{\includegraphics{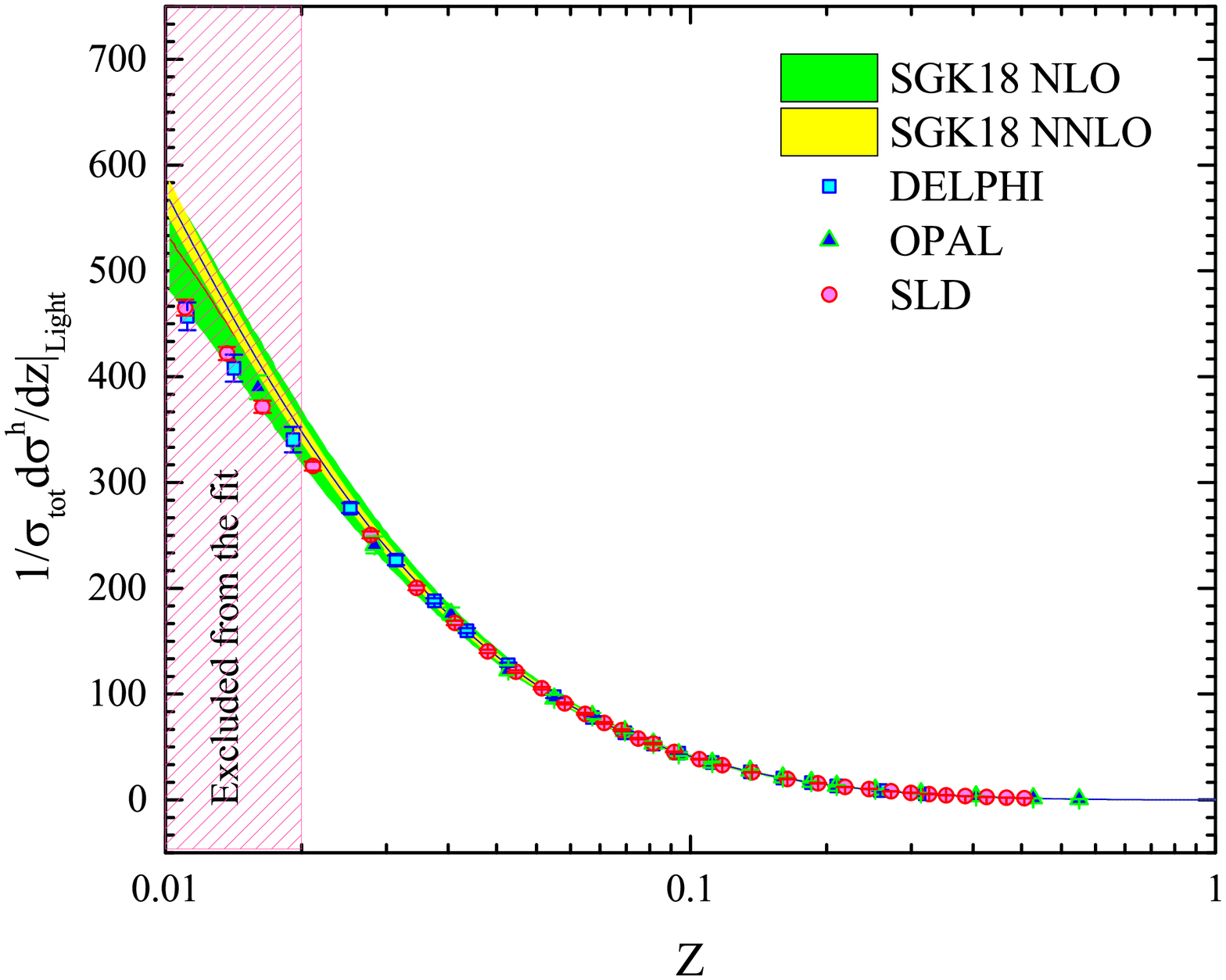}} 
		\resizebox{0.48\textwidth}{!}{\includegraphics{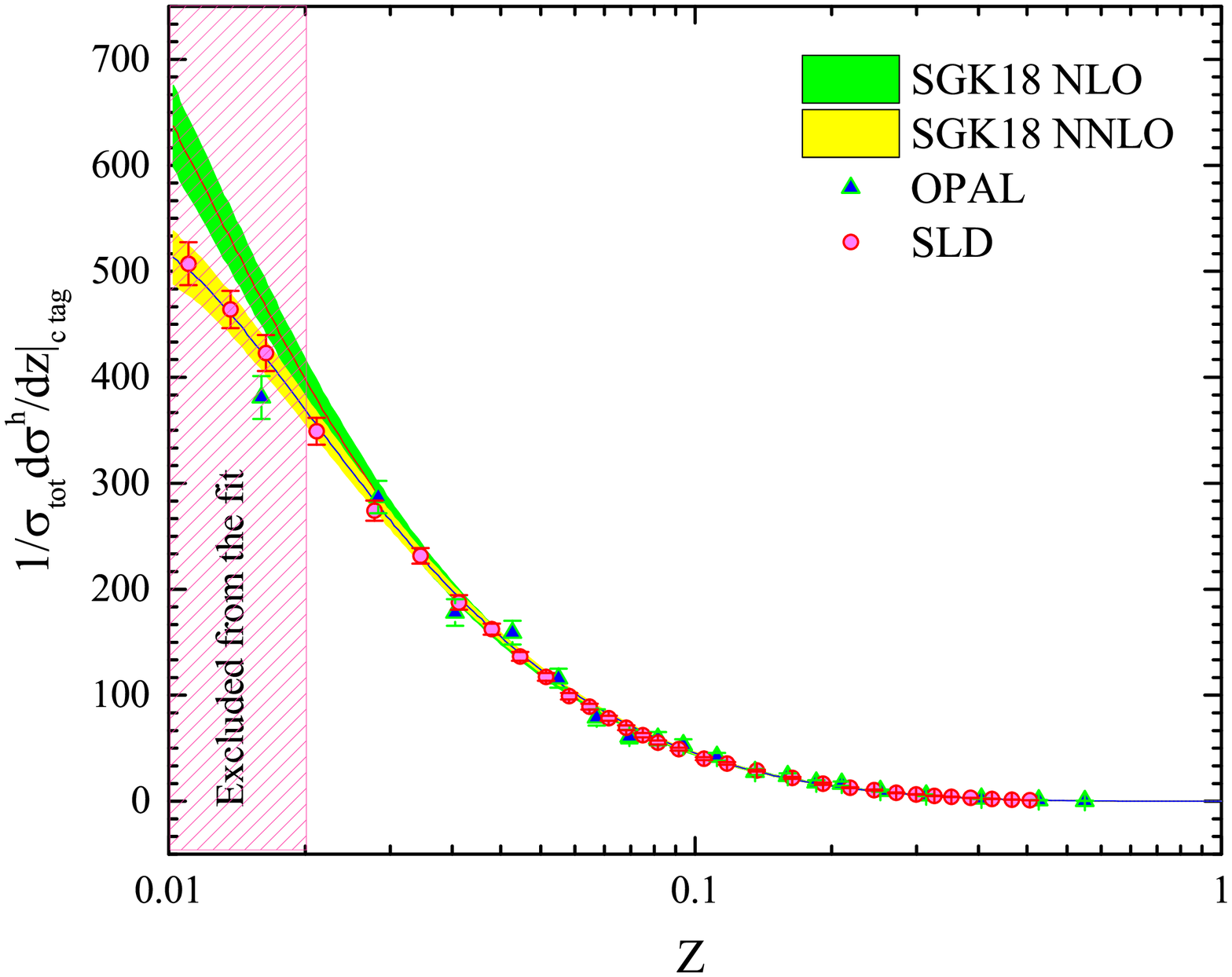}}  
		\resizebox{0.48\textwidth}{!}{\includegraphics{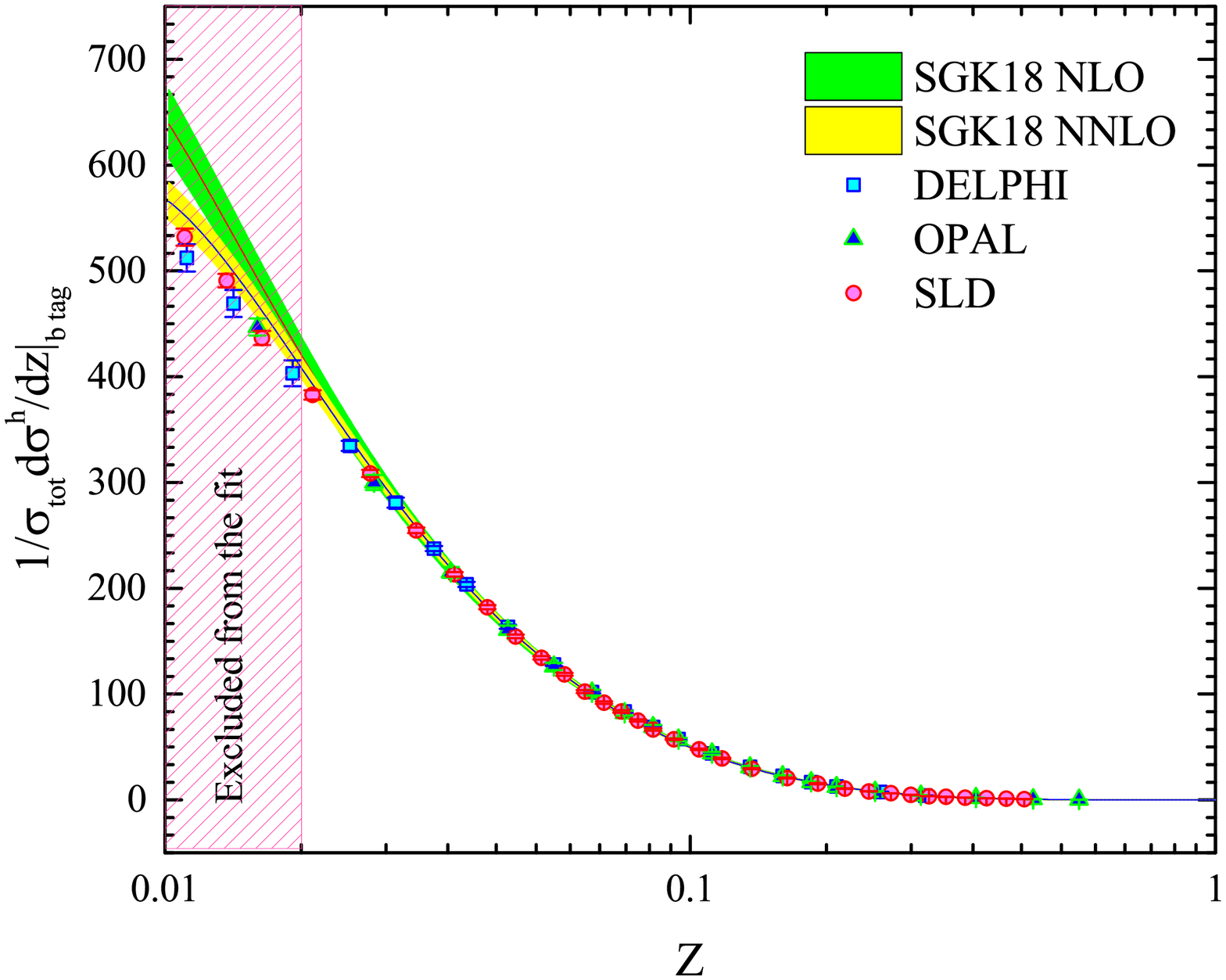}} 
		\caption{ ${\tt SGK18}$ NLO (solid line) and NNLO (dashed line) theory predictions for the normalized 
			total inclusive cross section, light, charm and bottom tagged ones of $D^h$-production compared with {\tt ALEPH}~\cite{Buskulic:1995aw}, {\tt OPAL}~\cite{Ackerstaff:1998hz,Akers:1995wt}, {\tt DELPHI}~\cite{Abreu:1998vq,Abreu:1997ir} and {\tt SLD}~\cite{Abe:2003iy} at the scale of $Q=M_Z$. 
			The shaded bands refer to our uncertainty results at NLO (green band) and NNLO (yellow band) and shaded areas
			indicate the kinematic regions excluded by our cuts. } \label{fig:cross1}
	\end{center}
\end{figure*}

\begin{figure*}[htb]
	\begin{center}
		\vspace{0.50cm}
		\resizebox{0.48\textwidth}{!}{\includegraphics{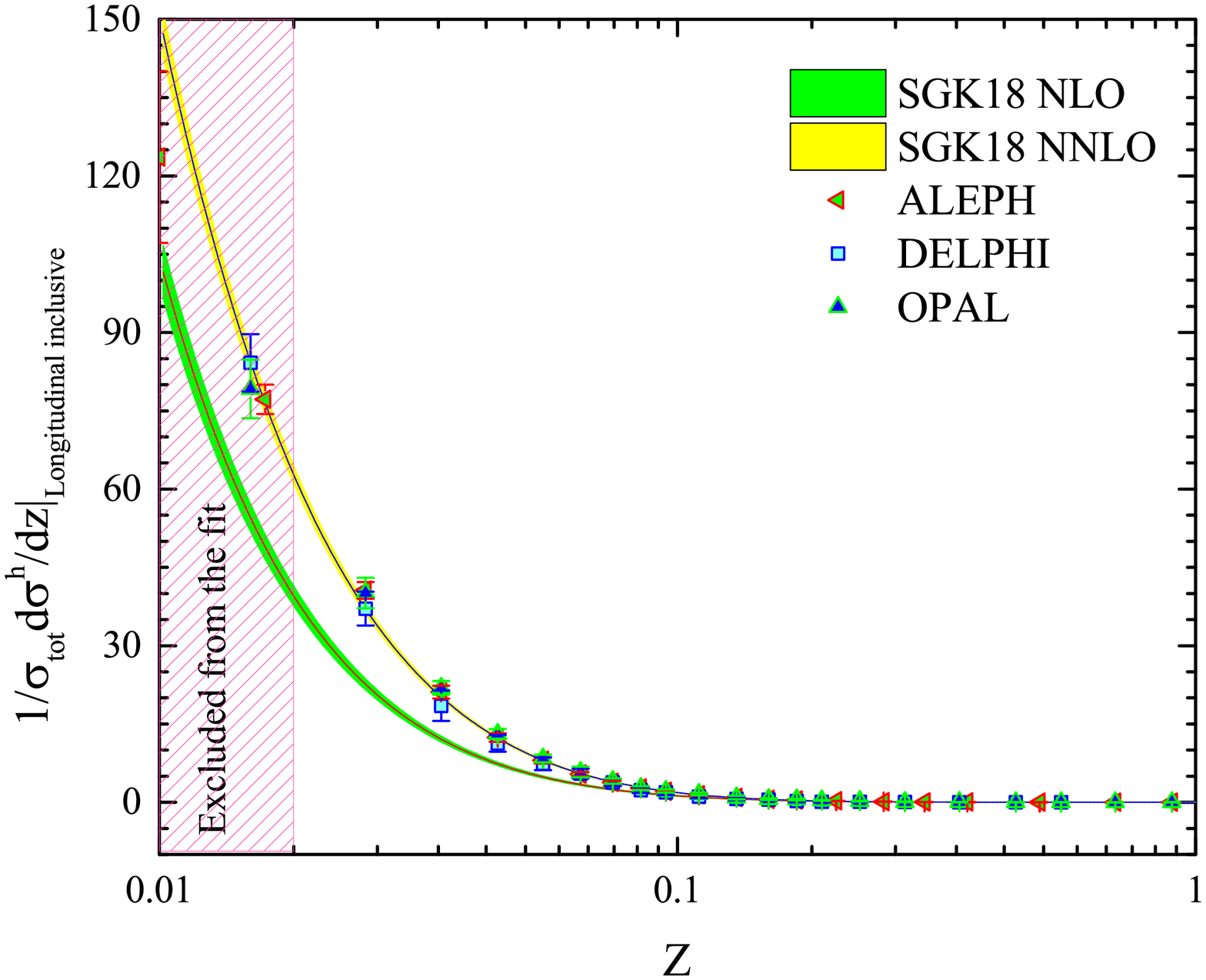}}  
		\resizebox{0.48\textwidth}{!}{\includegraphics{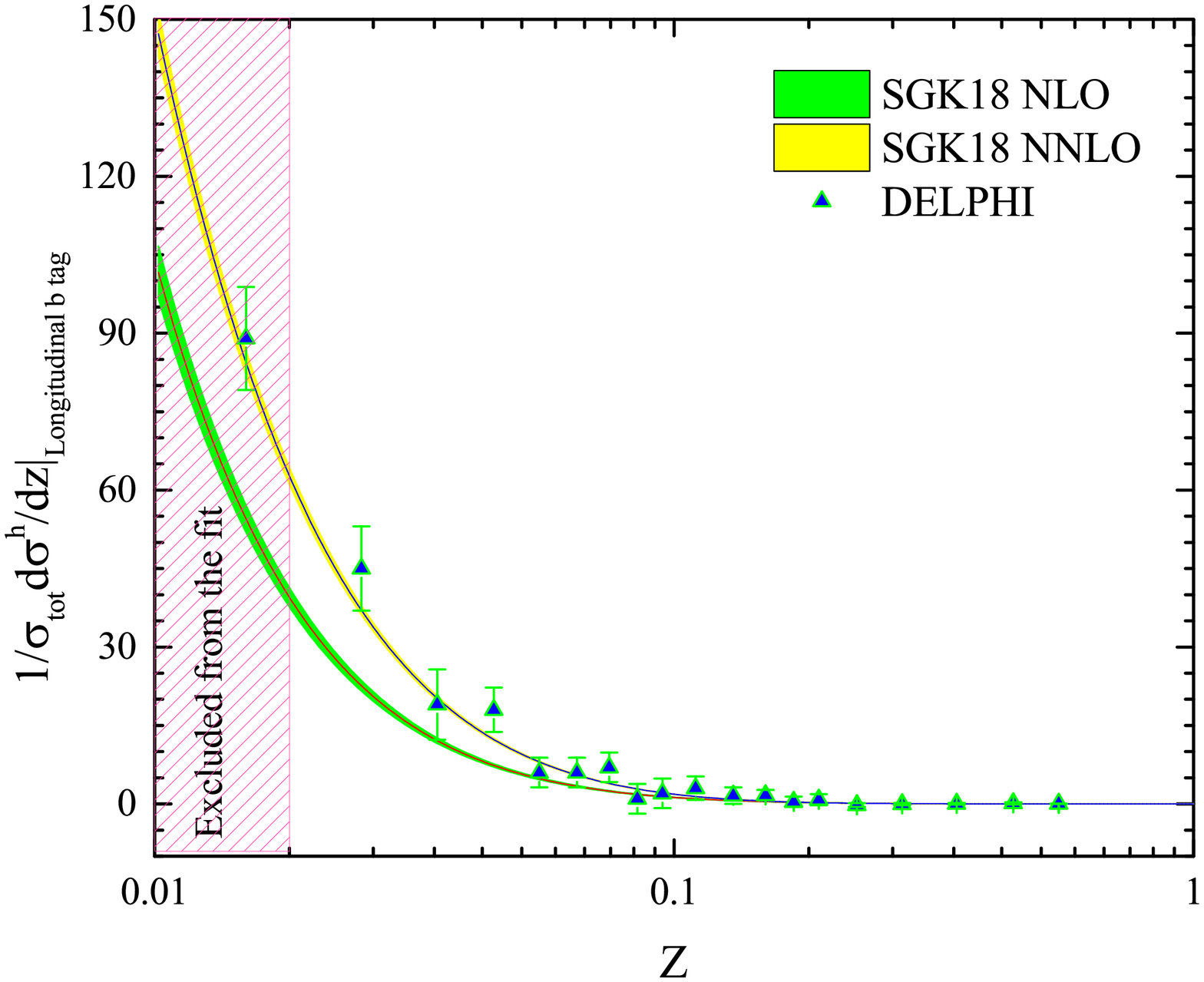}} 
		\caption{ Same as Fig.~\ref{fig:cross1} for the {\tt ALEPH}, {\tt DELPHI} and {\tt OPAL} longitudinal inclusive and longitudinal bottom tagged measurements. } \label{fig:crosslong}
	\end{center}
\end{figure*}

%
\subsection{ Discussion of {\tt SGK18} fit quality and data/theory comparison } 
%

In this section, we present the {\tt SGK18} NLO and NNLO theoretical predictions for the total and tagged SIA cross sections.
We also compare in details our results with all single-inclusive $D^h$ charged hadron production in $e^-e^+$ annihilation data analyzed in this study.
The values of the $\chi^2$ for both individual and total data sets included in {\tt SGK18} analysis have been reported in Table~\ref{tab:datasets} at both NLO and NNLO accuracy.
As one can see, the $\chi^2$ for inclusive and bottom tagged longitudinal data sets remarkably decrease at NNLO approximation. 

In order to judge the quality of the fits of {\tt SGK18} FFs analysis, we compare the experimental data to their corresponding NLO and NNLO theoretical predictions calculated using the NLO and NNLO FFs obtained from the QCD fits.
Fig.~\ref{fig:cross1} shows a comparison between the normalized total cross sections from the {\tt ALEPH}, {\tt DELPHI}, {\tt  OPAL} and {\tt  SLD} measurements of {\it unidentified} charged hadron and our NLO and NNLO predictions.
The uncertainty bands of the predictions, due to the one-$\sigma$ FF uncertainties, have also been shown in this figure. Moreover, the same comparisons have been performed for light ({\tt DELPHI}, {\tt OPAL} and {\tt SLD}); charm ({\tt OPAL} and {\tt SLD}); and bottom ({\tt DELPHI}, {\tt OPAL} and {\tt SLD}) tagged cross sections. Finally, we have shown the same comparison for the inclusive and $b$-tagged longitudinal cross sections from {\tt ALEPH}, {\tt DELPHI} and {\tt OPAL} data sets in Fig.~\ref{fig:crosslong}.   

Overall, the results obtained demonstrate a good agreement between the {\tt  SGK18} theoretical predictions and analyzed experimental data.
Considering the exclusion of small $z$ data points, the {\tt SGK18} results are also in reasonable agreement with data in the small and large $z$ regions for all data sets.
In Figs.~\ref{fig:cross1}, the NLO and NNLO predictions are in a satisfactory agreement in comparison to the total inclusive, light, charm and bottom tagged data for all range of $z$.

In order to investigate the fit quality of the total datasets at NLO and NNLO, as a next step, we discuss the size of uncertainty bands at NLO and NNLO. As one can see, the NLO uncertainty bands are slightly larger than NNLO one as presented in Fig.~\ref{fig:cross1}.
As it is seen from Fig.~\ref{fig:crosslong}, the ${\tt SGK18}$ NNLO theoretical predictions show more consistency with the data in comparison to the NLO ones for the inclusive and $b$-tagged longitudinal cross sections.
The NLO theoretical predictions tend to be overshoot by {\tt ALEPH}, {\tt DELPHI} and {\tt  OPAL} longitudinal experimental data for $z<0.1$ and {\tt DELPHI} $b$-tagged one for $z<0.2$.
The  data sets for the longitudinal inclusive and tagged cross sections have important effect on the determination of gluon FF, because they are non-vanishing already at LO ${\cal O}(\alpha _s)$ contribution.
According to the absence of precise data for wider range of $Q^2$, the longitudinal data could help to constrain the gluon FF. 

We should notice here that, in spite of the exclusion of small $z < 0.02$ data points, our NNLO theory predictions are in good agreement with the excluded region for $c$-tagged in Fig.~\ref{fig:cross1} and all other data sets in Fig.~\ref{fig:crosslong}.

\begin{figure*}[htb]
	\begin{center}
		\vspace{0.50cm}
		\resizebox{0.48\textwidth}{!}{\includegraphics{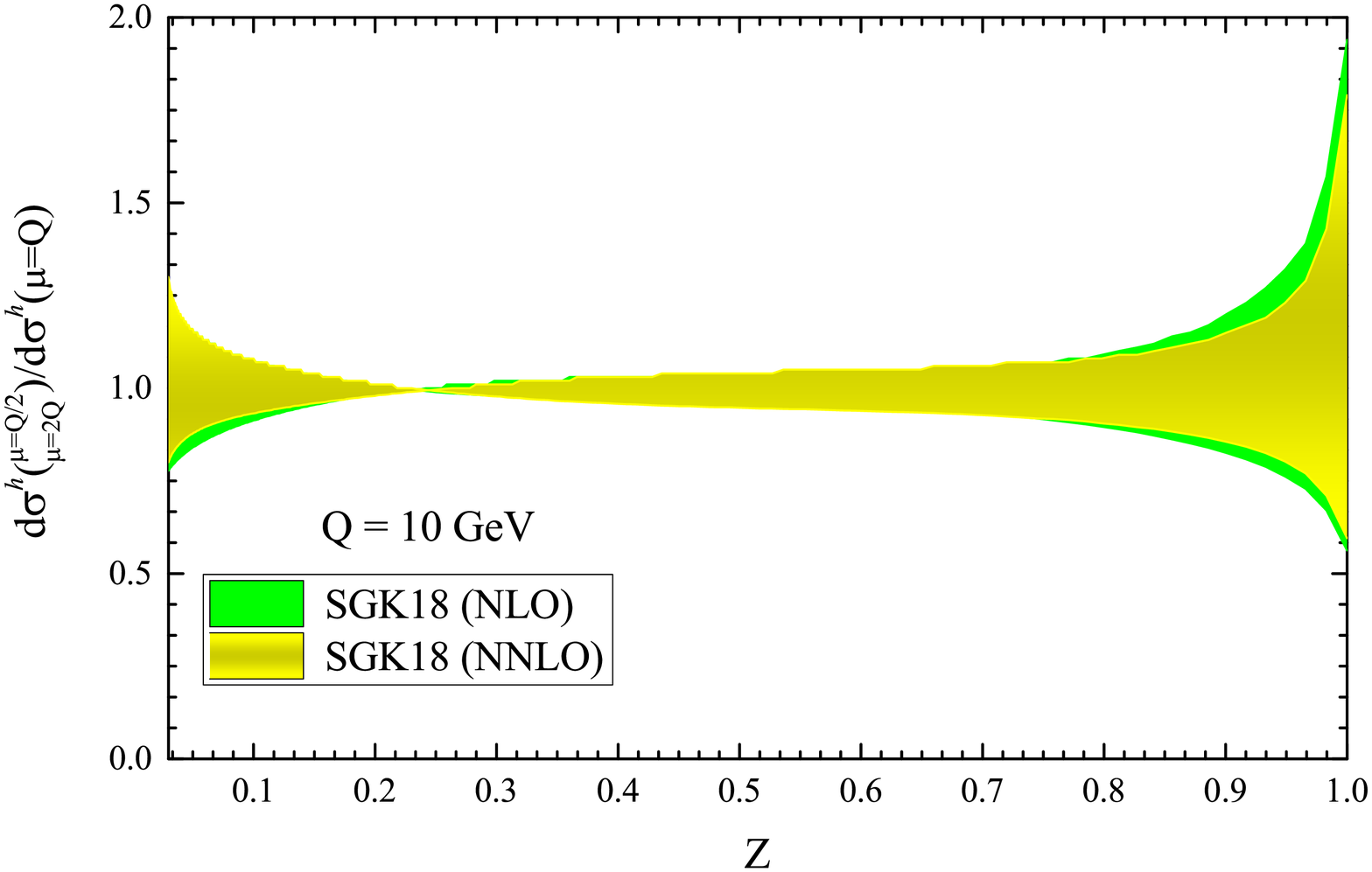}}  
		\resizebox{0.48\textwidth}{!}{\includegraphics{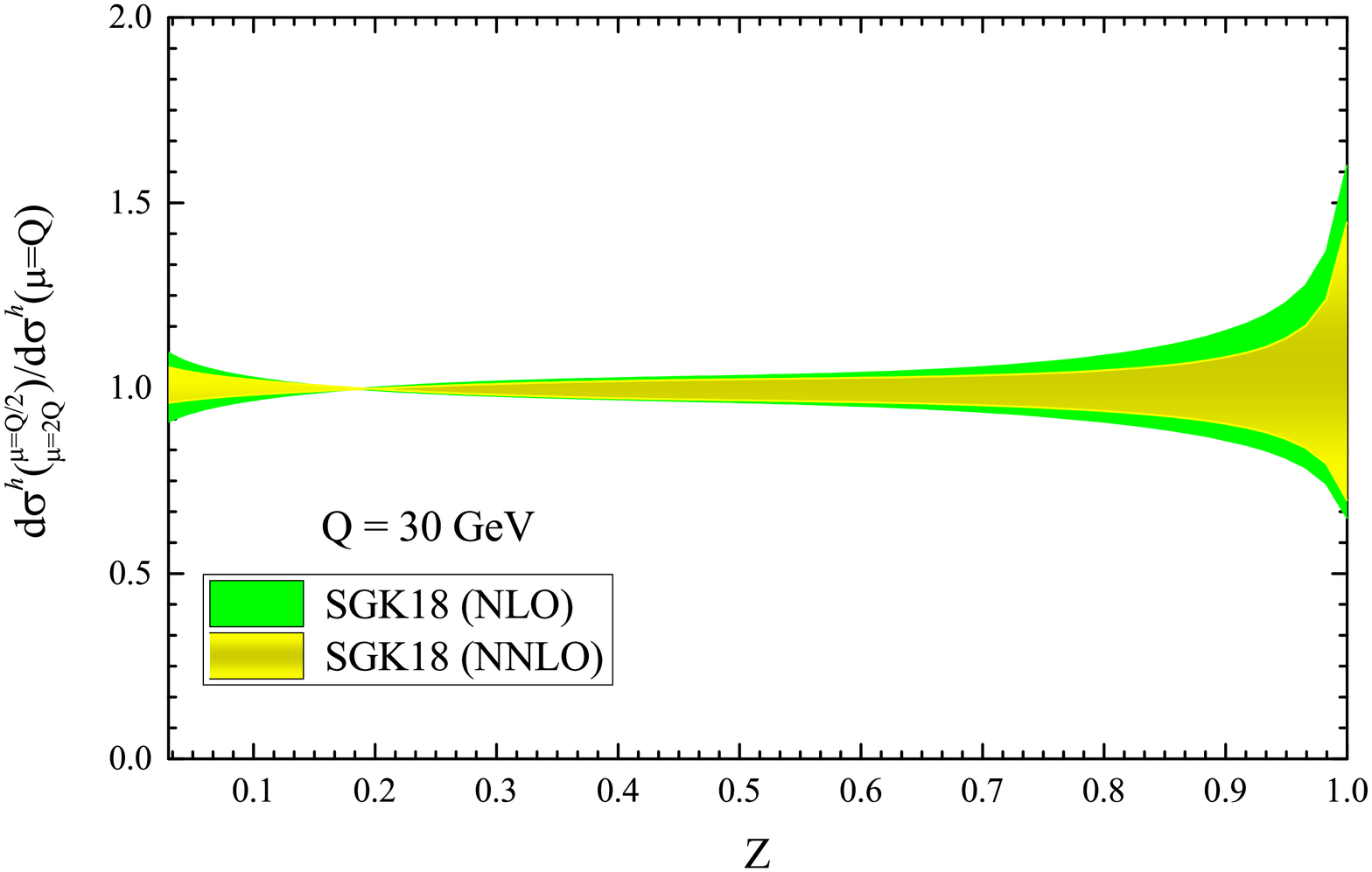}} 
		\resizebox{0.48\textwidth}{!}{\includegraphics{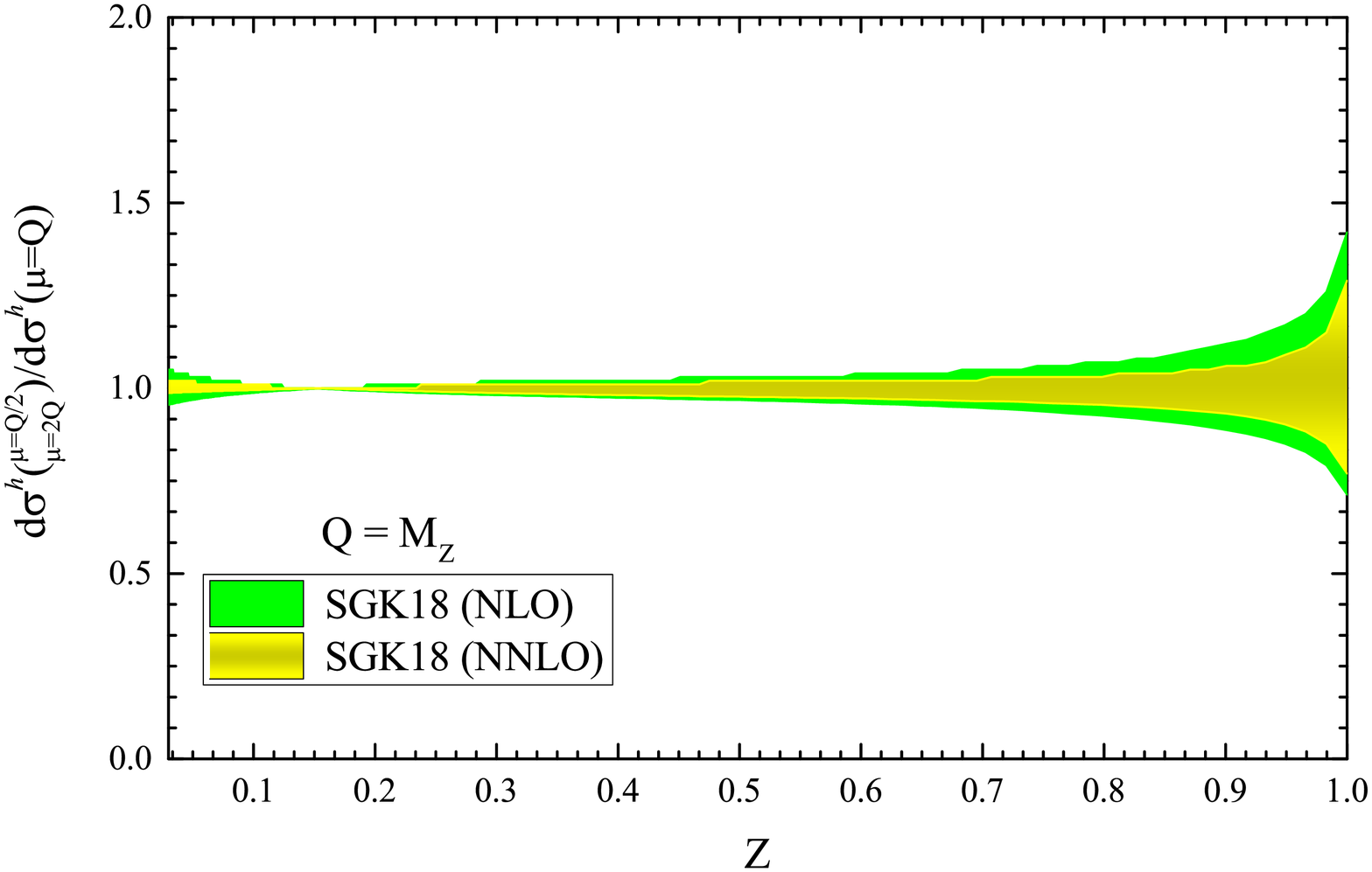}} 
		\caption{ Scale dependence of the SIA cross section at NLO and NNLO accuracy in the range of $Q/2 \le \mu \le 2Q$ normalized to the results obtained for $\mu = Q$
			for three values of $\sqrt{s}$.} \label{fig:Ratio}
	\end{center}
\end{figure*}

%
\subsection{ The improvement of NNLO accuracy in theoretical uncertainty} 
%

The fragmentation function uncertainties have different sources that classify into the experimental data errors and the theoretical uncertainties caused from the phenomenological assumptions in any global QCD fits.
The possible sources of theoretical uncertainties may include, for example, higher order correction effects in calculation of cross sections, the phenomenological form for the FFs parametrizations at an arbitrary initial scale and different assumptions of flavor and symmetries.

In this section, we present our results for studying the residual dependence on the choice of scale of energy $\mu$.
The most important source of theoretical uncertainties is dependence on the choice of the scale of energy $\mu$.
We expect to shrink progressively when we include higher and higher order corrections. It is exactly what we find in our study.

In Fig.~\ref{fig:Ratio}, the best fits of the NLO and NNLO analyses for {\it unidentified} charged hadron have been used to demonstrate the residual theoretical uncertainties due to the variations of the renormalization ($\mu _R$) and factorization ($\mu _F$) scales.
According to this figure, the SIA cross section depends on the scale of energy and the results have been shown at NLO and NNLO accuracies (shaded bands) for $\mu_R = \mu_F = \mu = Q/2$ and $\mu = 2 Q$ which are normalized to our default choice of $\mu = Q$.
It is obvious that the theoretical calculations depend on the scale of $\mu$. Note that our results have been presented for three scales of energy, $Q=10$~GeV, $Q=30$~GeV and $Q=M_Z$. According to the results presented in Fig.~\ref{fig:Ratio}, one can clearly conclude that the NNLO predictions are more stable than the NLO ones and come with much smaller theoretical uncertainties.

%
\section{Summary and Conclusions} \label{sec:conclusion}
%

In this paper, a new determination of {\it unidentified} charged hadron FFs at NLO and for the first time at NNLO accuracy in perturbative QCD are presented.
The flavor-untagged and the tagged SIA data in $e^- e^+$ annihilation are included in this analysis that are reported by CERN ({\tt ALEPH}~\cite{Buskulic:1995aw}, {\tt OPAL}~\cite{Ackerstaff:1998hz,Akers:1995wt}, and {\tt DELPHI}~\cite{Abreu:1998vq,Abreu:1997ir}), SLAC ({\tt TPC}~\cite{Aihara:1988su} and {\tt SLD}~\cite{Abe:2003iy}) and DESY ({\tt TASSO}~\cite{Braunschweig:1990yd}). The heavy flavor contributions are considered in the ZM-VFNS in $z$-space in the framework of publicly available {\tt APFEL} code.

We illustrate the quality of the {\tt SGK18} FFs at NLO and NNLO and show that the results presented in this analysis are in good agreement with the results in literature and all exprimental data analyzed in this study. We have presented the uncertainties for the $D^h$ fragmentation functions and the corresponding theory predictions using the ``Hessian'' approach.  

The most striking remarkable improvements to emerge from {\tt SGK18} FFs analysis are as follows:
As a first improvement, this study is the first step towards enhancing our understanding of parton-to-{\it unidentified} charged hadrons FFs by analyzing flavor-untagged and the tagged SIA data considering the NNLO accuracy in perturbative QCD.
As a second improvement, we use smoother kinematical cut for the small $z$ regions than other analyses in the literature such as {\tt DSS07}. Consequently, {\tt SGK18} FFs analysis uses a wider range of exprimental data points in the fitting procedure.

As a third improvement, we have presented the perturbative stable QCD fits and observed a reduction of uncertainties for our FFs as well as theory predictions at NNLO with respect to NLO. Finally, we have chosen $Q_0 = 5$~GeV as an initial scale in {\tt SGK18} fits and then the number of active flavor is always fixed to $n_f=5$. This choice improve the fit because time-like matching conditions are unknown at NNLO. Within this choice of initial scale, the heavy quark threshold as well as the matching condition don't need to be taken into account.

As a final improvement, by using our fit result at NNLO for FFs, the agreement between our predictions for the inclusive and $b$-tagged longitudinal cross sections have improved in comparison with the NLO analysis which suggest that the inclusion of higher order corrections could improve the fit quality. 

We hope that our research will serve as a base for future studies on the determination of {\it unidentified} charged hadrons FFs from wide range of exprimental observables at CERN, HERA and SLAC. However further works need to be carried out to establish a framework to consider the SIDIS and hadron collider data into the analysis.

A {\tt FORTRAN} package, which evaluates the {\tt SGK18} NLO and NNLO {\it unidentified} charged hadron FFs as well as the theory predictions presented in this study can be obtained from the authors upon request via electronic mail.

%
\begin{acknowledgments}
%

Authors thank School of Particles and Accelerators, Institute for Research in Fundamental Sciences (IPM) for financial support of this project.
Hamzeh Khanpour also is thankful the University of Science and Technology of Mazandaran for financial support provided for this research.

\end{acknowledgments}
%



\end{document}